\documentclass[12pt]{article}

\usepackage{amsmath,amssymb,collref,graphicx,subfigure}
\usepackage{hyperref}
\usepackage{cite}

\graphicspath{{/}
              {Figures/}
              }

\newcommand{\dd}{d}
\newcommand{\ii}{i}

\newcommand{\Ux}{\ensuremath{U(1)_x}}
\newcommand{\mz}{\ensuremath{m_x}}
\newcommand{\ma}{\ensuremath{m_{A^x}}}
\newcommand{\tz}{\ensuremath{\tan\zeta}}
\newcommand{\mx}{\ensuremath{M_x}}
\newcommand{\lagrange}[1]{\mathcal{#1}}
\newcommand{\order}[1]{\ensuremath{\mathcal{O}(#1)}}
\newcommand{\abs}[1]{\left\lvert #1 \right\rvert}
\newcommand{\expect}[1]{\left\langle #1 \right\rangle}
\newcommand{\beq}{\begin{equation}}
\newcommand{\eeq}{\end{equation}}
\newcommand{\bea}{\begin{eqnarray}}
\newcommand{\eea}{\end{eqnarray}}
\newcommand{\lrf}[2]{\left(\frac{#1}{#2}\right)}
\newcommand{\met}{\ensuremath{\,/\hspace{-0.25cm}E_T}}

\newcommand{\nnmb}{\nonumber}
\newcommand{\del}{\partial}
\newcommand{\tev}{\, {\rm TeV}}
\newcommand{\gev}{\, {\rm GeV}}

\newcommand{\ccdot}{\!\cdot\!}
\newcommand{\tmp}{\tilde{\mathcal{P}}}

\begin{document}

\begin{titlepage}
\noindent
\vspace{1cm}
\begin{center}
  \begin{Large}
    \begin{bf}
LHC Signatures of a Minimal\\
Supersymmetric Hidden Valley
     \end{bf}
  \end{Large}
\end{center}
\vspace{0.2cm}

\begin{center}

\begin{large}
Yuk Fung Chan$^{(a,c)}$, Matthew Low$^{(b,c)}$,\\
David E. Morrissey$^{(c)}$, and Andrew P. Spray$^{(c)}$
\end{large}
\vspace{1cm}\\
  \begin{it}
(a) Program in Applied and Computational Mathematics, Princeton University, Princeton, NJ 08544, USA\\
(b) Enrico Fermi Institute, University of Chicago, 5640 South Ellis Avenue, Chicago, IL 60637, USA\\
(c) TRIUMF, 4004 Wesbrook Mall, Vancouver, BC V6T 2A3, Canada
%
\vspace{0.2cm}\\
email: dmorri@triumf.ca, aps37@triumf.ca
\vspace{0.2cm}
\end{it}

\end{center}

\center{\today}


\begin{abstract}

We investigate the LHC signals of a minimal supersymmetric
hidden valley.  Our theory consists of the supersymmetric Standard
Model along with a light hidden \Ux\ gauge multiplet and a pair
of hidden chiral superfields that spontaneously break the new Abelian
gauge symmetry near a GeV\@.  The visible and hidden sectors
interact exclusively through supersymmetric gauge kinetic mixing.
We perform a thorough examination of the hidden decay cascades initiated
by the lightest Standard Model superpartner and we study the range
of LHC signals they can produce.  In particular, we find parameter
regions that give rise to missing energy, single and multiple lepton jets,
and displaced vertices.  Given the simplicity of the underlying theory
and the broad range of collider signals it can produce, we propose
that this model is a useful benchmark for LHC studies of (supersymmetric)
hidden valleys.

\end{abstract}

\end{titlepage}

\setcounter{page}{2}


\section{Introduction\label{sec:intro}}

  The Large Hadron Collider~(LHC) is rapidly collecting
data from high energy proton ($pp$) collisions with a center of
mass energy of $\sqrt{s} = 7\,\tev$.  Precision measurements
of collision events by the ATLAS and CMS detectors may lead to the
discovery of new physical phenomena beyond those predicted by the
Standard Model~(SM) of particle physics.  In order to maximize the
experimental reach for such new discoveries, it is necessary to
search broadly within the data.  For this, it is helpful to have
a diverse set of theoretical benchmarks to compare against
observations~\cite{Alves:2011wf,Morrissey:2009tf}.

  Hidden valley~(HV) models are an interesting and well-motivated
class of new physics that could be discovered at the LHC\@.
However, compared to many other theories of exotic particle
physics phenomena, their signatures at the LHC have been
studied in much less detail.  HV models can also be challenging
to find at the LHC, leading to events that are not picked up
by the more standard searches for new physics~\cite{Strassler:2006im,
Strassler:2006qa,Strassler:2006ri,Han:2007ae}.
For these reasons, in the present work we present and investigate
a minimal supersymmetric HV theory that can serve as a benchmark
for future collider searches.

  A generic HV theory consists of three components~\cite{Strassler:2006im}:
a visible sector of particles containing the SM, a hidden sector consisting
of particles that couple feebly to the SM, and some mediator states
that couple efficiently to both the visible and the hidden sectors.
Furthermore, the hidden sector is typically assumed to have a characteristic
mass well below the electroweak scale while the mediators have
TeV-scale masses.\footnote{All of these mass scales can be natural
in the context of supersymmetry~\cite{ArkaniHamed:2008qn,Cheung:2009qd,Katz:2009qq,ArkaniHamed:2008qp,Morrissey:2009ur}
or strong coupling/warping~\cite{Strassler:2006im,
Juknevich:2009ji,McDonald:2010iq,Gherghetta:2010cq}.}
The presence of mediators coupling to the SM allows for the abundant
production of hidden particles at high-energy colliders.
When a mediator is created, it can decay to light HV states
and populate the valley.
In turn, the light hidden particles can be stable, or more interestingly
they might be able to decay back to the SM\@.  This broad paradigm
of new physics scenarios can give rise to an equally broad range
of collider signals, including missing energy~($\met$), displaced vertices,
and high particle multiplicities.

  Low-scale supersymmetry with $R$-parity provides a particularly
promising and effective mediator to a hidden valley.
Once produced, SM superpartners will decay in a cascade down
to the lightest SM superpartner~(LSMP).  However, if there are lighter
superpartners among the hidden states, the LSMP will be unstable against
decaying into one of them.  The decay cascade would then continue within
the hidden valley until the true lightest superpartner~(LSP), now derived
from the hidden sector, is produced.  Along the way, several hidden
particles decaying back to the SM could also be generated.
The net collider signature of such an event would therefore consist
of the direct products of the supersymmetric SM cascade,
together with missing energy from the LSP, and the products
of the HV decays.

  In this paper we investigate the collider signals of a simple
and minimal supersymmetric hidden valley.  The theory consists of the minimal
supersymmetric standard model~(MSSM) together with a much lighter
hidden sector consisting of a supersymmetric \Ux\ gauge multiplet
and a pair of chiral multiplets $H$ and $H'$ that spontaneously break
the new gauge symmetry at an energy scale well below the electroweak.
The MSSM and hidden sectors are assumed to couple
exclusively through a gauge kinetic mixing interaction~\cite{Holdom:1985ag}.

  Despite its minimality, this theory gives rise to a rich variety of
collider signatures.  As such, it provides a simple but useful
benchmark model to guide LHC searches for hidden valleys.
Even so, we emphasize that this model does not come close to
covering the full range of possible collider signals that can
emerge from the HV paradigm.
Related studies of supersymmetric hidden valleys have appeared
in Refs.~\cite{ArkaniHamed:2008qp,Baumgart:2009tn,Cheung:2009su,Bai:2009it,
Carloni:2010tw,Carloni:2011kk}.
Relative to these previous works,
we focus on a single specific hidden sector theory for which we
perform a detailed investigation of the possible hidden decay cascades,
we take into account the entire structure of the hidden sector,
and we attempt to characterize the full range of collider signals
that it can produce.

  The outline of this paper is as follows.  In Sec.~\ref{sec:model}
we present our benchmark hidden valley in detail,
while in Sec.~\ref{sec:feature} we investigate some its general
features that are relevant for LHC phenomenology.  We perform a scan
over the parameters characterizing the HV in Sec.~\ref{sec:scan}
to classify the range of hidden cascade topologies.
In Sec.~\ref{sec:bench} we study the LHC signals of five specific
sample parameter points.  Finally, Sec.~\ref{sec:conc} is
reserved for our conclusions.  Some additional technical details
are contained in Appendices~\ref{appa}, \ref{appb} and \ref{appc}.

\bigskip

\section{Model\label{sec:model}}

  Our model is the minimal extension of the MSSM by a new spontaneously
broken \Ux\ gauge group.  In this sense, it is also a minimal
supersymmetric hidden valley.
We review the pertinent features of the model here;
for more details see Ref.~\cite{Morrissey:2009ur}.

\subsection{Fields and Interactions}

  The hidden sector consists of the exotic \Ux\ vector superfield $X$
and two chiral superfields $H$ and $H'$ responsible for breaking the
hidden gauge symmetry.
Both are neutral under the SM gauge symmetries but have \Ux\
charges $\pm 1$; while all MSSM superfields are neutral under \Ux.
The superpotential is extended by a hidden mass term
\begin{equation}
  W = W_{MSSM} - \mu' \, H \, H' .
\end{equation}
The K\"ahler potential is taken to be canonical, while the gauge kinetic terms
include mixing between $X$ and the hypercharge superfield $B$:
\begin{align}
  \lagrange{L} & \supset \frac{1}{4} \int d^2 \theta \, \bigl( X^\alpha X_\alpha + 2 \epsilon \, B^\alpha X_\alpha + B^\alpha B_\alpha \bigr) + h.c. \notag \\
  & \supset \epsilon \biggl( - \frac{1}{2} B_{\mu\nu} X^{\mu\nu} + \frac{\ii}{2} \, \tilde{B}^\dagger \bar{\sigma} \cdot \partial \tilde{X} + \frac{\ii}{2} \tilde{X}^\dagger \bar{\sigma} \cdot \partial \tilde{B} + D_Y D_X \biggr) . \label{eq:kmixfield}
\end{align}
This term can be naturally induced by integrating out heavy fields charged
under both hypercharge and \Ux, in which case values of
$\epsilon \sim 10^{-4}$\,--\,$10^{-2}$ are typical.  Note that in terms of
component fields, we have kinetic mixing of the gauge fields and the gauginos,
as well as $D$-term mixing that will contribute to the scalar potential.

  We will be largely agnostic about the specific method of SUSY breaking,
assuming only that the hidden sector soft masses are real and less than
or on the order of a GeV\@.  An explicit mechanism that realizes this
spectrum of soft masses in a natural way is gauge mediation by
gauge messengers charged under the MSSM but neutral under the hidden \Ux.
This leads to hidden sector soft parameters on the order
$\epsilon \, m_{MSSM} \lesssim\,\gev$.  To avoid a cosmologically
troublesome light fermion in this scenario~\cite{Morrissey:2009ur},
additional GeV-scale supergravity contributions
are needed, and these also provide a natural origin for the $\mu'$ term
through the Giudice-Masiero mechanism~\cite{Giudice:1988yz}.
As such, in this mechanism the soft terms of the hidden sector are not
calculable without an underlying theory of supergravity, and we treat them as
free parameters.  The relevant supersymmetry-breaking operators are\footnote{
We neglect a possible mass term mixing $\tilde{X}$ and $\tilde{B}$.
No such term is generated from gauge mediation to the MSSM
alone~\cite{Morrissey:2009ur}, and the residual gravity-mediated
contribution is on the order of $\epsilon\,M_x \ll M_x$.}
\begin{equation}
  - \lagrange{L}_{\text{hid, soft}} = m_H^2 \abs{H}^2 + m_{H'}^2 \abs{H'}^2
+ \biggl( - b^{\prime}\, H \, H' + \frac{1}{2} M_x \tilde{X} \tilde{X}
+  h.c. \biggr) ,
\end{equation}
where we may take $b^{\prime}$ and $M_x$ positive without loss of generality.

  The cost of making $b'$ and $M_x$ both real and positive is that
there can remain a physical CP-violating phase in $\mu'$.  As mentioned above,
we will assume that $\mu'$ is real (but not necessarily positive) as well.
This is done largely for convenience, since a full exploration of
the effects of hidden-sector CP violation lies beyond the scope
of this paper.  However, will comment on the qualitative implications on 
the mass spectrum and mixing matrices in Sec.~\ref{subsec:mix}, 
and on the decay modes in Sec.~\ref{sec:feature}.


  Electroweak symmetry breaking in the MSSM introduces additional
contributions to the hidden sector scalar potential.  The MSSM Higgs
vacuum expectation values~(VEVs) induce a non-vanishing hypercharge $D$-term
\begin{equation}
  \expect{D_Y} \equiv \xi_Y = - \frac{g_Y}{2} \cos 2\beta \, v^2 .
\end{equation}
The $D_Y D_X$ term in \eqref{eq:kmixfield} then leads to an effective
Fayet-Iliopoulos term~\cite{Fayet:1974jb} for the hidden sector.
This term can be absorbed by shifting the hidden
Higgs soft masses according to
\begin{equation}
  m_H^2 \to \tilde{m}_H^2 = m_H^2 - \epsilon \, g_x \xi_Y \,,
\quad m_{H'}^2 \to \tilde{m}_{H'}^2 = m_{H'}^2 + \epsilon \, g_x \xi_Y \,.
\end{equation}
Note that this contribution would break supersymmetry in the hidden sector even
if it were isolated from all other sources of breaking.

\subsection{Hidden Sector Masses and Mixings \label{subsec:mix}}

  We demand that the \Ux\ gauge symmetry is spontaneously broken by
a non-zero VEV of $H$ or $H'$.
Since the hidden scalar potential is structurally identical to that
of the neutral MSSM Higgses (at tree-level),
both hidden scalar expectation values can be taken positive,
\begin{equation}
  \expect{H} = \eta \, \sin\zeta \,, \qquad \expect{H'} = \eta \, \cos\zeta \,,
\end{equation}
with $\zeta \in [0, \pi/2]$.  In terms of the underlying Lagrangian parameters, we have
(neglecting $\order{\epsilon^2}$ corrections)
\begin{align}
\sin2\zeta & = \frac{2b^\prime}{2|\mu^\prime|^2
+\tilde{m}_H^2+\tilde{m}_{H^\prime}^2} \,,
\label{alfa}\\
\eta^2 & = -\frac{|\mu^\prime|^2}{g_x^2}
+\frac{-\tilde{m}_H^2\tan^2\zeta
+\tilde{m}_{H^\prime}^2}{g_x^2(\tan^2\zeta-1)} \,.
\label{eta}
\end{align}
As a result of the symmetry breaking, the hidden gauge boson
acquires a mass of
\begin{equation}
  \mz = \sqrt{2} \, g_x \eta \,.
\end{equation}
These results apply for vanishing kinetic mixing.  We will show below
that the modifications to these expressions from the mixing are negligibly
small.

  After \Ux\ breaking, there are seven physical states in the hidden sector:
the gauge boson $X_\mu$, two real scalars $h_{1,2}^x$ and a pseudoscalar $A^x$,
and three Majorana fermions $\chi_j^x$.  The hidden sector introduces
seven new parameters compared to the MSSM, which we take to be
$\epsilon$, $g_x$, $\mu'$, \mx, \mz, \tz\ and the pseudoscalar mass \ma.

  We turn next to the gauge kinetic mixing.  The neutral gauge boson quadratic terms are
\begin{multline}
  \lagrange{L} \supset - \frac{1}{4} X^{\mu\nu} X_{\mu\nu} - \frac{1}{4} A^{\mu\nu} A_{\mu\nu} - \frac{1}{4} Z^{\mu\nu} Z_{\mu\nu} - \frac{\epsilon}{2} X^{\mu\nu} \bigl( c_W A_{\mu\nu} - s_W Z_{\mu\nu} \bigr) \\
  + \frac{1}{2} \mz^2 X^\mu X_\mu + \frac{1}{2} m_Z^2 Z^\mu Z_\mu \,,
\end{multline}
where $s_W \, (c_W) = \sin \theta_W \, (\cos \theta_W)$.  To linear order in $\epsilon$, we can remove the kinetic mixing
by the transformations\footnote{It is straightforward
to generalise these expressions to all orders in $\epsilon$.}
\begin{equation}
  \begin{split}
    X_\mu & \to X_\mu + \epsilon \, s_W Z_\mu \,, \\
    Z_\mu & \to Z_\mu \,, \\
    A_\mu & \to A_\mu - \epsilon \, c_W X_\mu \,.
  \end{split}
\label{eq:vecrot}\end{equation}
This rotation introduces a mass mixing between $X_\mu$ and $Z_\mu$.
However, the mixing is suppressed by both $\epsilon$ and $\mz^2/m_Z^2$,
and so can be safely ignored.  The rotation also induces a coupling of
the $X$ vector boson to the MSSM with effective charge
$-\epsilon\,c_W(e/g_X)$, and a coupling of the $Z$ boson to all the
hidden scalars and fermions with strength proportional to $\epsilon\,s_W$.

  The Bino-Xino kinetic mixing can be most conveniently removed by
the field rotation (again dropping terms \order{\epsilon^2})
\begin{equation}
  \begin{split}
    \tilde{X} & \to \tilde{X} - \epsilon \tilde{B} \,, \\
    \tilde{B} & \to \tilde{B} \,.
  \end{split}
\label{eq:binoxino}\end{equation}
This introduces a mass mixing between the hidden and visible
neutralinos.  The resultant mixing angle is less than or on the order
of $\epsilon m_{\chi_1^x}/m_{\chi^0_1}$.  Since this angle is expected
to be very small, we diagonalize the hidden and visible neutralino
mass matrices independently, and treat the mixing in the mass insertion approximation in the few cases where it is relevant.

  The four visible sector neutralinos simply mix as in the MSSM\@.
The mass matrix for the three hidden sector fermions
is given in the basis
$\tilde{\psi}^x \equiv (\tilde{H},\tilde{H'},\tilde{X})^t$ by
\beq
\mathcal{M}_X = \begin{pmatrix}
  0 & -\mu' & \mz s_{\zeta}\\
  -\mu' & 0 & -\mz c_{\zeta}\\
  \mz s_{\zeta} & -\mz c_{\zeta} & M_x\\
\end{pmatrix}
 .
\eeq
We relate the gauge and mass eigenstates by an analogue of the
SLHA convention for the MSSM~\cite{Skands:2003cj}. Explicitly,
\begin{equation}
  {\psi}_j^x = P^\ast_{ij} \chi_i^x \,,
\end{equation}
where $P$ is a unitary matrix.  As all hidden soft parameters are
assumed to be real, we take $P$ to have strictly real entries at the
cost of allowing some of the $\chi_j^x$ mass eigenvalues to be negative,
which we sort by
$\abs{m_{\chi_1^x}} \leq \abs{m_{\chi_2^x}} \leq \abs{m_{\chi_3^x}}$.

  Among the scalars, there is no kinetic mixing between between the
visible and hidden sectors.  However, there is mass mixing among the
CP-even states induced by the $D$-term mixing.  Since the resultant
mixing angles are very small, on the order of $\epsilon\,m_x/m_Z \ll 1$,
it is a good approximation to diagonalize the hidden and visible scalar
sector mass matrices independently and treat the mixing as a mass
insertion interaction when it is needed.

  In the hidden sector this produces two CP-even states
$h_{1}^x$ and $h_{2}^x$ and a CP-odd ${A^x}$.
The hidden pseudoscalar ${A^x}$ mass is
\begin{equation}
\ma^2 = \frac{2b^\prime}{\sin 2\zeta} = 2|\mu^\prime |^2
+\tilde{m}_H^2+\tilde{m}_{H^\prime}^2\,.
\label{amass}\end{equation}
For the two CP-even states, $h_{1}^x$ and $h_{2}^x$,
the mass matrix is given by
\bea
\mathcal{M}^2_{h^x} =
\begin{pmatrix}
\mz^2 s_{\zeta}^2+\ma^2c_{\zeta}^2 & -(\mz^2+\ma^2)s_{\zeta}c_{\zeta}
\\
-(\mz^2+\ma^2)s_{\zeta}c_{\zeta} & \mz^2c_{\zeta}^2 + \ma^2s_{\zeta}^2
\end{pmatrix}
.
\label{hmass}\eea
Due to the structural similarity to the MSSM,
it follows that the lighter mass eigenstate ${h_1^x}$ has mass bounded
from above by both $\mz \abs{\cos 2\zeta}$ and $\ma \abs{\cos 2\zeta}$.
The heavier CP-even scalar is in turn bounded from below,
as $m_{{h_1^x}}^2 + m_{{h_2^x}}^2 = \mz^2 + \ma^2$.  The expansion of
$H, H'$ in terms of the mass basis is
\begin{equation}
  \begin{split}
    H & = \eta\,\sin\zeta + \frac{1}{\sqrt{2}} \bigl(
R_{1a}h^x_a
+ \ii\,\cos \zeta\, {A^x} \bigr) \,, \\
    H' & = \eta\,\cos\zeta
+ \frac{1}{\sqrt{2}} \bigl(
R_{2a}h^x_a
+ \ii\,\sin \zeta\,{A^x} \bigr) \,,
  \end{split}
\end{equation}
where $R$ is the orthogonal matrix that diagonalizes the real symmetric matrix in Eq.~\eqref{hmass}.

Let us comment upon the modifications to these results when we allow the hidden sector parameters to be complex.  The fact that we can rotate away any phase associated with $b'$ means that at tree level all our results for the scalars are unchanged.  In particular, the hidden Higgses can not spontaneously break CP.  We would expect loop corrections to mix $A^x$ with $h_{1,2}^x$, but this should not significantly alter the spectrum or $h_1^x$--$h_2^x$ mixing.  In the fermion sector, we can no longer choose the mixing matrix $P$ to be real, but there are no qualitatively new effects.

  For completeness, we collect a list of the most important interaction
terms within the hidden sector in Appendix~\ref{appa}.  These interactions are
expressed in terms of both gauge and mass eigenstates, and are valid both when the hidden sector parameters are real and (at tree-level) complex.

\bigskip

\section{To and From the Hidden Valley\label{sec:feature}}

  Hidden sector states can be produced directly through their induced
couplings to charged SM particles~\cite{Essig:2009nc,Reece:2009un,
Bjorken:2009mm,Batell:2009yf}, from rare decays of
the $Z$~\cite{Baumgart:2009tn,Cheung:2009su},
and at the end of MSSM cascades~\cite{Strassler:2006qa,ArkaniHamed:2008qp}.
The rates for the first two of these channels are suppressed by at
least a factor of $\epsilon^2$,
but no such suppression is present for production via the MSSM\@.
MSSM superpartner production can therefore be the dominant
source of hidden states at high energy colliders, and we will
focus exclusively on it in the present work.

  With $R$-parity, $R$-odd MSSM superpartners are produced in pairs
with full gauge strength and decay in a cascade down to the lightest MSSM
superpartner.  This would-be LSP is unstable, and can decay to the
lighter hidden states through the kinetic mixing coupling.
The decay cascade continues in the hidden sector, possibly producing some
$R$-even hidden states along the way.  In turn, these can decay back
to the SM through kinetic mixing interactions.  The fact that we can both enter
and leave the hidden sector through the same supersymmetric operator
is a particular elegance of this model.

  In this section we discuss the production of hidden states from MSSM
decay cascades along with decays from the hidden sector back to the SM\@.
We also classify the general structure of supersymmetric events at the
Tevatron and the LHC in the presence of the minimal light hidden sector.

\subsection{(MSSM) Decays Into the Hidden Sector}

  All of the MSSM superpartners can decay into the hidden sector.
However, the partial widths are proportional to $\epsilon^2$,
and can usually be ignored for all fields other than the lightest SM
superpartner~(LSMP), which has no other decays open to it.
The most important coupling for LSMP decays generally comes from
the \Ux\ gaugino interactions:
\begin{align}
  - \lagrange{L} & \supset \sqrt{2} \, g_x \bigl( H^\ast \tilde{H} \tilde{X} - H'{}^\ast \tilde{H}' \tilde{X} \bigr) \notag \\
  & \to \sqrt{2} \, g_x \bigl( H^\ast \tilde{H} \tilde{X} - H'{}^\ast \tilde{H}' \tilde{X} \bigr) - \epsilon \sqrt{2} \, g_x \bigl( H^\ast \tilde{H} \tilde{B} - H'{}^\ast \tilde{H}' \tilde{B} \bigr) . \label{eq:decaytohid}
\end{align}
Note that since the LSMP is unstable,  the usual dark matter constraints
do not apply and in particular, it need not be a neutralino.
This divides supersymmetric hidden valley theories into four
subclasses depending on the nature of the LSMP: a gluino, a sfermion, 
a neutralino, or an approximately degenerate chargino-neutralino pair (the structure of the MSSM mass matrix makes a LSMP chargino difficult to obtain~\cite{Gherghetta:1999sw}).

 A gluino LSMP decays primarily through a four-body mode involving
an off-shell squark and an off-shell neutralino, or in a three-body
mode with an off-shell squark that relies on hidden-MSSM neutralino
mass mixing.  In both cases, two energetic quark jets are produced
together with one or two hidden states.  These decays can be prompt
or significantly displaced, depending on the spectrum.
For $m_{{\chi}^0} \sim m_{\tilde{q}} \gg m_{LSMP}$
there is a very strong kinematic suppression that can lead to
a long-lived gluino and $R$-hadron-like
signals~\cite{Baer:1998pg,Hewett:2004nw,Fairbairn:2006gg}.

  The case of a sfermion LSMP was considered in Ref.~\cite{Baumgart:2009tn}.
It can decay in a three-body mode to the corresponding fermion and a pair
of hidden states via an off-shell Bino, or in a two-body channel from
mass mixing (induced by kinetic mixing) between the visible and
hidden neutralinos. Both channels can be comparable in rate,
and are typically prompt unless there is additional suppression
from kinematics or a very small value of the kinetic mixing.
In particular, the kinematic suppression becomes severe if the
sfermion is much lighter than the MSSM neutralinos~\cite{Baumgart:2009tn}.

A light chargino has some similarities to a sfermion; it can decay into the hidden sector to $W$/Higgs plus either two hidden states through an off-shell Bino, or one through the neutralino mass mixing.  The chargino can also usually decay to the LSMP neutralino plus a pion~\cite{Gherghetta:1999sw}; since the pion is soft, it usually escapes detection and so this has the same collider signal as for the neutralino.  All three chargino decays are prompt, and which dominates depends on the mass splitting with the neutralino.

  In the present work we will focus on the third case
where the LSMP is a neutralino.
The interaction of Eq.~\eqref{eq:decaytohid} induces direct
two-body decays to a hidden sector fermion and scalar (or vector).
The total associated decay width is
\begin{align}
  \Gamma & = \frac{1}{8\pi} g_x^2 \epsilon^2 |N_{11}^*|^2m_{LSMP} \\
  & \simeq (2\times 10^{-18}\text{s})^{-1}\, {|N_{11}|^2} \lrf{g_x}{0.3}^2 \lrf{\epsilon}{10^{-3}}^2 \lrf{m_{LSMP}}{100\,\gev} , \notag
\end{align}
where $|N_{11}|$ is the Bino fraction of the LSMP neutralino.
These decays are prompt for parameter values remotely close
to the fiducial values quoted here.  

  Decays of a neutralino LSMP to a hidden neutralino and a scalar
($h_1^x,$ $h_2^x,$ $A^x$) arise in an obvious way from the interactions
of Eq.~\eqref{eq:decaytohid}.  Decays to the vector are also expected
from the Goldstone equivalence principle~\cite{Baumgart:2009tn},
although their origin is a bit murkier in the unitary gauge
(which we use here) where the Goldstone mode does not appear explicitly.
In this gauge, the vector decay mode
arises from the small mass mixing between the Bino and the hidden Higgsinos
induced by the interaction of Eq.~\eqref{eq:decaytohid}.
The corresponding mixing angle goes like
$(\epsilon\,\mz/m_{\chi^0_1}) \ll \epsilon$,
but the mass suppression in this angle is cancelled in the squared matrix
element by the kinematic enhancement of the coupling to the longitudinal
mode of the vector that arises for $\mz \ll m_{\chi^0_1}$,
as expected from the Goldstone boson equivalence
theorem~\cite{Cornwall:1974km,Lee:1977eg}.

  The Goldstone equivalence theorem also implies that the decays of the
LSMP neutralino should populate the scalar and vector channels
equally, up to kinematic effects.  Explicitly, we find (neglecting
the kinematic effects of the hidden masses) the decay widths
\begin{equation}
  \Gamma (\chi_1^0 \to \chi^x_j S^x) =
\epsilon^2\frac{g_x^2 m_{\chi_1^0}|N_{11}|^2}{32 \pi} \times
  \begin{cases}
    \abs{s_\zeta P_{j1}^\ast + c_\zeta P_{j2}^\ast}^2 & \text{if } S^x = X_{\mu} \,, \\
    \abs{c_\zeta P_{j1}^\ast - s_\zeta P_{j2}^\ast}^2 & \text{if } S^x = {A^x} \,, \\
    \abs{R_{11} P_{j1}^\ast + R_{12} P_{j2}^\ast}^2 & \text{if } S^x = {h_1^x} \,, \\
    \abs{R_{12} P_{j1}^\ast - R_{11} P_{j2}^\ast}^2 & \text{if } S^x = {h_2^x} \,.
  \end{cases}
\label{eq:lsmpdecay}\end{equation}
Taking $S^x = {A^x}$ (for example) and summing over the hidden
fermion flavours, we find
\begin{align}
  \sum_j \Gamma (\chi_1^0 \to \chi^x_j {A^x}) & =
\epsilon^2\frac{g_x^2 m_{\chi_1^0}|N_{11}|^2}{32 \pi}
\nnmb\\
&~~~~~~~\times\,
\sum_j \Bigl[ c_\zeta^2 P_{j1}^\ast P_{j1} + s_\zeta^2 P_{j2}^\ast P_{j2}
- 2 s_\zeta c_\zeta Re(P_{j1}^\ast P_{j2} \bigr) \Bigr] \notag \\
  & = \epsilon^2\frac{g_x^2 m_{\chi_1^0}|N_{11}|^2}{32 \pi} \ , \label{eq:lsmpscalar}
\end{align}
where we have used the unitarity of $P$ in the last line.
It is clear that the same result holds for the CP-even scalars as well
as the vector, and even holds when the hidden sector breaks CP.  In contrast, the relative decay widths to the hidden
neutralinos are not democratic.  It is clear from
Eq.~\eqref{eq:decaytohid} that the mostly-Higgsino states
will be preferred; summing over the scalar channels
in Eq.~\eqref{eq:lsmpdecay} and using orthogonality of $R$, we find that
\begin{equation}
  \sum_{S^x} \Gamma (\chi_1^0 \to \chi^x_j {S^x}) = \epsilon^2 \frac{g_x^2 m_{\chi_1^0}|N_{11}|^2}{16 \pi} \bigl( \abs{P_{j1}}^2 + \abs{P_{j2}}^2 \bigr) \,.
\label{eq:lsmptofermion}\end{equation}
In particular, in the limit that the mass and gauge bases overlap,
the LSMP will decay to the two Higgsinos with equal probability
(and the gaugino not at all).

\subsection{Decays Out of the Hidden Sector\label{subsec:decayout}}

All of the hidden states are electrically neutral and uncoloured,
and are not directly visible at the LHC\@.  Nevertheless,
hidden particles can contribute more than just missing energy
to collider events if they decay back to the SM\@.
The only applicable coupling to induce such a decay is the gauge kinetic mixing.
We show here that this can allow the vector $X_{\mu}$ and the lightest
scalar $h_{1}^x$ to decay primarily to the SM to produce new observable
effects in collider detectors, while the remaining hidden
states decay almost entirely to other (possibly off-shell) hidden states.

  Starting with the vector $X_{\mu}$, its decays to the SM proceed
at the rate~\cite{Batell:2009yf,Reece:2009un}
\begin{equation}
\Gamma(X\to SM) = \Gamma(X\to~hadrons) + \sum_{\ell=e,\mu,\tau}\Gamma(X\to\ell\bar{\ell}) \label{eq:xdecay}
\end{equation}
where
\begin{align}
\Gamma(X\to \ell\bar{\ell}) & = \frac{\alpha\,\epsilon^2\,c_W^2}{3}
m_x\sqrt{1-\frac{4m_{\ell}^2}{m_x^2}}\left(1+\frac{2m_{\ell}^2}{m_x^2}\right) \notag \\
&\simeq (1.1\times 10^{-5}\text{cm})^{-1}\lrf{\epsilon}{10^{-3}}^2
\lrf{m_x}{1\,\gev} ,
\end{align}
and
\begin{equation}
\Gamma(X\to~hadrons) = \Gamma(X\to\mu^+{\mu}^-)\,R(s=m_x^2) \,,
\label{eq:xwidthhad}\end{equation}
with $R(s) = \sigma(e^+e^-\to hadrons)/\sigma(e^+e^-\to\mu^+{\mu}^-)$,
measured values of which are tabulated in Ref.~\cite{Nakamura:2010zzi}.
These decay modes typically only dominate the width of $X_{\mu}$
when all the possible two-body hidden decay channels,
$X_\mu \to {A^x} {h_1^x}$ and $X_\mu \to \chi^x_i \chi^x_j$,
are kinematically forbidden.
On the other hand, if the vector can only decay to the SM,
it is apt to be a rich source of additional
fermions in supersymmetric collider events.

  The lightest hidden scalar ${h_1^x}$ can decay to the SM
through its gauge-induced coupling to the vector $X_{\mu}$
and its mass mixing with the CP-even MSSM Higgs bosons.
Both contributions to the total width are small relative to the mass,
and usually relevant only when all the possible two-body hidden
decays, $h_{1x}\to \chi^x_i\chi^x_j$, are kinematically forbidden.
Even if the $h_{1}^x$ mode decays exclusively to the SM,
its lifetime can be so long that it decays far outside a typical
collider detector and contributes only missing energy to collider events.

  The long lifetime of ${h_1^x}$ in the minimal model
under consideration arises because it is constrained by the structure of
the hidden sector to be lighter than the $X_{\mu}$ vector.
As such, the only vector-induced decays of this light scalar are a
four-body mode with two off-shell vectors, and a loop-mediated decay
containing the vector~\cite{Batell:2009yf}.
In both cases the corresponding decay rates are proportional to $\epsilon^4$,
and are each too slow to allow the ${h_1^x}$ to decay within the LHC
(or Tevatron) detectors.

\begin{figure}[ttt]
\begin{center}
  \includegraphics[width=0.6\textwidth]{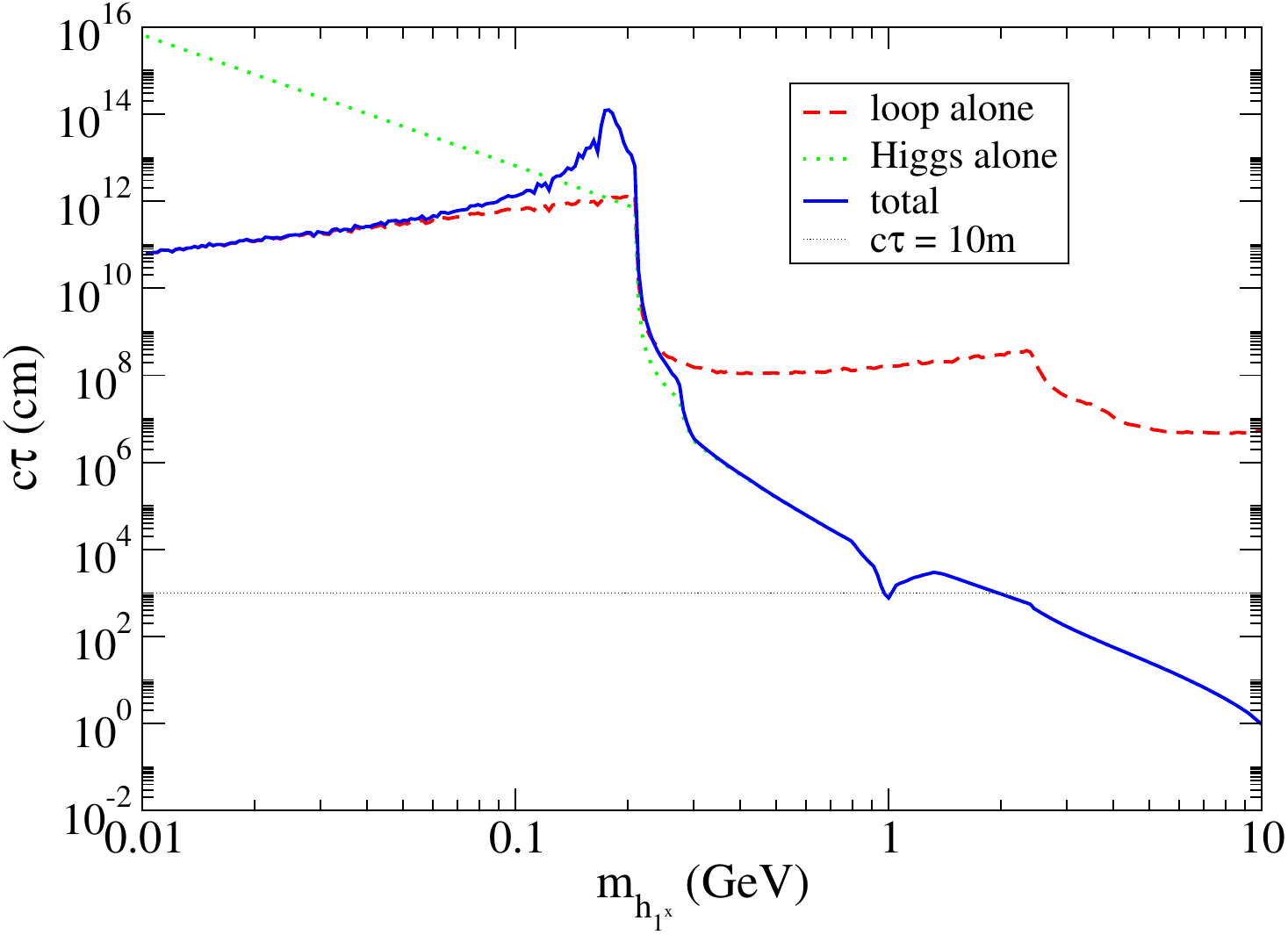}
\end{center}
\caption{Decay length ($c\tau$) of the lightest hidden Higgs boson $h_{1}^x$
for different decay mechanisms.  In making this figure we have
fixed $\epsilon=10^{-3}$, $m_x = 1.5\,m_{h_1^x}$,
$c^2_{2\beta}(s_{\zeta}R_{1a}-c_{\zeta}R_{2a})^2 = 1$, and $m_{h^0} = 125\,\gev$.
The horizontal dotted line corresponds to $c\tau=10\,\text{m}$,
roughly the radius of the LHC detectors.
}
\label{fig:h1xdecay}
\end{figure}

  Mixing with the MSSM Higgs bosons by way of $D$-term interactions (listed
in Appendix~\ref{appa}, see Eq.~(\ref{eq:higgsmix})) also contributes to the
decay width of $h_1^x$.
The inclusive decay rate produced by this mixing is
(in the MSSM Higgs decoupling limit)
\begin{align}
  \Gamma_{h_{1}^x} & = \epsilon^2s_W^2\lrf{m_xm_Z}{m_{h^0}^2}^2
  c^2_{2\beta}\,(s_{\zeta}R_{1a}-c_{\zeta}R_{2a})^2\Gamma_{h^0}(m_{h_1^x})  \\
  & \simeq (7.8\times 10^{-12})\,\Gamma_{h^0}(m_{{h_1^x}}) \notag \\
  & \quad \times\,c^2_{2\beta}\,(s_{\zeta}R_{1a}-c_{\zeta}R_{2a})^2\,
  \lrf{\epsilon}{10^{-3}}^2\lrf{m_x}{\gev}^2\lrf{125\,\gev}{m_{h^0}}^4 ,
  \nnmb
\end{align}
where $\Gamma_{h^0}(m_{{h_1^x}})$ is the decay width of a SM Higgs
boson with mass equal to that of $h_1^x$.  For ${h_1^x}$ masses
above the two-muon threshold, these Higgs-mixing decays dominate over
those induced by the hidden vector.  We illustrate the relative
values of the decay length $c\tau$ for both decay mechanisms
(considered individually) in Fig.~\ref{fig:h1xdecay}
for $\epsilon=10^{-3}$, $m_x = 1.5\,m_{h_1^x}$,
$c_{2\beta}^2(s_{\zeta}R_{1a}-c_{\zeta}R_{2a})^2 = 1$, and $m_{h^0} = 125\,\gev$.  Note that the partial width to light hadrons includes the effects of hadronic form factors from Ref.~\cite{Donoghue:1990xh}, relevant in the region $2m_\pi \lesssim m_{h_1^x} \lesssim 1$~GeV.  The total width accounts for  interference between the two contributions, which depends on the absolute sign of $\epsilon$.
From this figure, we see that the $h_{1}^x$ state is stable on
collider time scales for $m_{h_1^x} \lesssim 1\,\gev$.
In the present work we will focus on a range of hidden sector parameters
for which the $h_1^x$ state is very long-lived, either due to a small mass
or from additional suppression due to mixing or kinematics.
However, for larger masses the decays of $h_{1}^x$ to heavy flavours may be
visible in the LHC detector, a signal that is similar to those
discussed for hidden valley models in Ref.~\cite{Strassler:2008fv}.

  The heavier hidden scalars typically decay much more efficiently within
the hidden sector than to the SM\@.  They can still contribute to visible
collider objects if they decay to one or two vectors:
\beq
{A^x} \to {h_1^x} X_\mu ; \quad {h_2^x} \to 2 X_\mu ; \quad h_2^x \to A^x X_\mu .
\eeq
Note that in the last decay one of the final states is always virtual.
These states may also decay to hidden neutralinos or multiple light scalars
(for $h_2^x$).  In all cases, the decays are typically prompt unless the
mass spectrum forces them to go far off-shell.

  Had we allowed violation of CP in the hidden sector, several new bosonic
decays could become possible.  Of these, two are potentially relevant:
\begin{equation}
  A^x \to X_\mu X_\nu ; \quad A^x \to h_1^x h^1_x \ .
\end{equation}
These decays are suppressed compared to the CP-conserving ones by 
the small $A^x$--$h_{1,2}^x$ mixing, but can be important in particular 
regions of parameter space.  The former decay is enhanced when 
$m_{A^x} \gg m_x$ by the coupling to the longitudinal mode of the vector.  
It is unlikely to be relevant for the parameter space we consider, 
since we assume no large mass hierarchies in the hidden sector.  
The decay to two scalars can be relevant when no other hidden sector 
decays of $A^x$ are open.  In this case, the leading mode in the 
CP-conserving case is to the SM through an off-shell vector.  
Whether the CP-violating decay can dominate this depends on 
the precise spectrum and the size of the $A^x$--$h_2^x$ mixing, 
which are model dependent qualities.  It is a model-independent statement 
that this decay mode will lead to fewer decays back to the 
visible sector (and thus more missing energy).

  By $R$-parity, the fermion decay modes all have the form
\beq
\chi_i^x \to \chi_j^x{S^x}^{(*)} \quad (i>j) \,,
\label{eq:fermiondecays}\eeq
where $S^x = {h_1^x},\,{h_2^x},\,{A^x},\,X_{\mu}$, and the bosonic
product may be off-shell.  In particular, if the $\chi_1^x \, h_1^x$ final state is forbidden, the dominant decay mode will normally be through an off-shell vector.  Calculating the width for this process requires an integration over the hidden vector width:
\begin{multline}
  \Gamma = \frac{\abs{G_{ij}}^2 \abs{m_{\chi_i^x}}}{16 \pi} \int \dd Q^2 \,
\frac{m_{\chi_i^x}^2}{Q^2}\,  \frac{\sqrt{Q^2} \, 
\Gamma (X \to SM; m_x = \sqrt{Q^2})}{(Q^2 - m_x^2)^2 + m_x^2 \Gamma_x^2} \\
  \times \sqrt{I ( \rho_1, \rho_2 )} \biggl[ (1 - \rho_1^2)^2 + \rho_2^2 (1 + \rho_1^2) - 2 \rho_2^4 + 6 \rho_2^2 \frac{\Re (G_{ij}^2)}{\abs{G_{ij}}^2} \biggr] ,
\label{eq:oshelldec}\end{multline}
where 
\begin{equation}
  \rho_1 = \frac{m_{\chi_j^x}}{m_{\chi_i^x}}, \quad \rho_2 = \frac{\sqrt{Q^2}}{m_{\chi_i^x}} ,
\end{equation}
the coupling $G_{ij}$ is
\begin{equation}
  G_{ij} = g_x (P_{i1} P_{j1}^\ast - P_{i2} P_{j2}^\ast ), 
\end{equation}
and the phase space function
\begin{equation}
  I (a, b) = 1 - 2a^2 - 2b^2 + a^4 + b^4 - 2a^2 b^2 .
\end{equation}
Note that unless the fermion mass splitting is less than $2m_\pi$, this requires an integration over the experimentally-measured function $R(s)$, defined below Eq.~\eqref{eq:xwidthhad}.  If the fermion mass splitting is less than $2m_e$, the decay will instead proceed through an off-shell $h_1^x$ to $\chi_1^x \gamma\gamma$.  As for the heavier fermions, these decays are prompt unless they are limited by kinematics.  The presence of CP violation would change the branching fractions for these modes, but lead to no new decays.

  In summary, the only significant exit channels from the hidden sector
are through the vector $X_{\mu}$ or the lightest scalar $h_1^x$,
with the latter occurring too late to be seen in typical colliders
for the parameter ranges to be considered in the present work.
Hidden vectors can be created directly from the LSMP decay, as well as
from cascade decays within the hidden sector.  Additional vectors
can also be produced from hidden showering as discussed
in Refs.~\cite{Bai:2009it,Cheung:2009su,Carloni:2010tw,Carloni:2011kk}.
Once a hidden vector is created, it will
decay back to the SM provided it has no open channels within
the hidden sector.  Indeed, this condition is close to a necessary
one for the hidden sector to contribute more than just missing
energy in high-energy collisions.

\subsection{Collider Events}

  Putting together the pieces from the discussion above, a simple
picture emerges for supersymmetric events at high-energy colliders.
MSSM superpartners are produced in pairs and decay in a cascade down to
the LSMP.  These decay subsequently to a hidden fermion
and a hidden boson.  The cascade continues within the
hidden sector, and some hidden vectors may be produced along the way.
If the vector decays back to the SM, the MSSM cascade will be accompanied
by additional SM particles as well missing energy from the stable
or long-lived hidden states.

  Since the MSSM LSMP is assumed to be much heavier
than any of the states in the hidden sector, the two decay
products of the LSMP will each be highly boosted by an amount
$\gamma \sim m_{LSMP}/m_{hid}$.  On each branch,
the cascade will continue, and any further decay products will be
collimated by an amount $\Delta R \sim 1/\gamma$.
In particular, a vector created along the way will produce a
highly collimated hadronic jet, or a \emph{lepton jet}
consisting of two or more collimated leptons with possible
additional hadronic activity~\cite{ArkaniHamed:2008qp,Baumgart:2009tn,
Cheung:2009su}.  The branching ratios of the vector follow from
Eq.~\eqref{eq:xdecay}.  We will refer to these highly collimated
objects together as HV jets.

\begin{figure}[ttt]
\begin{center}
  \includegraphics[width=0.5\textwidth]{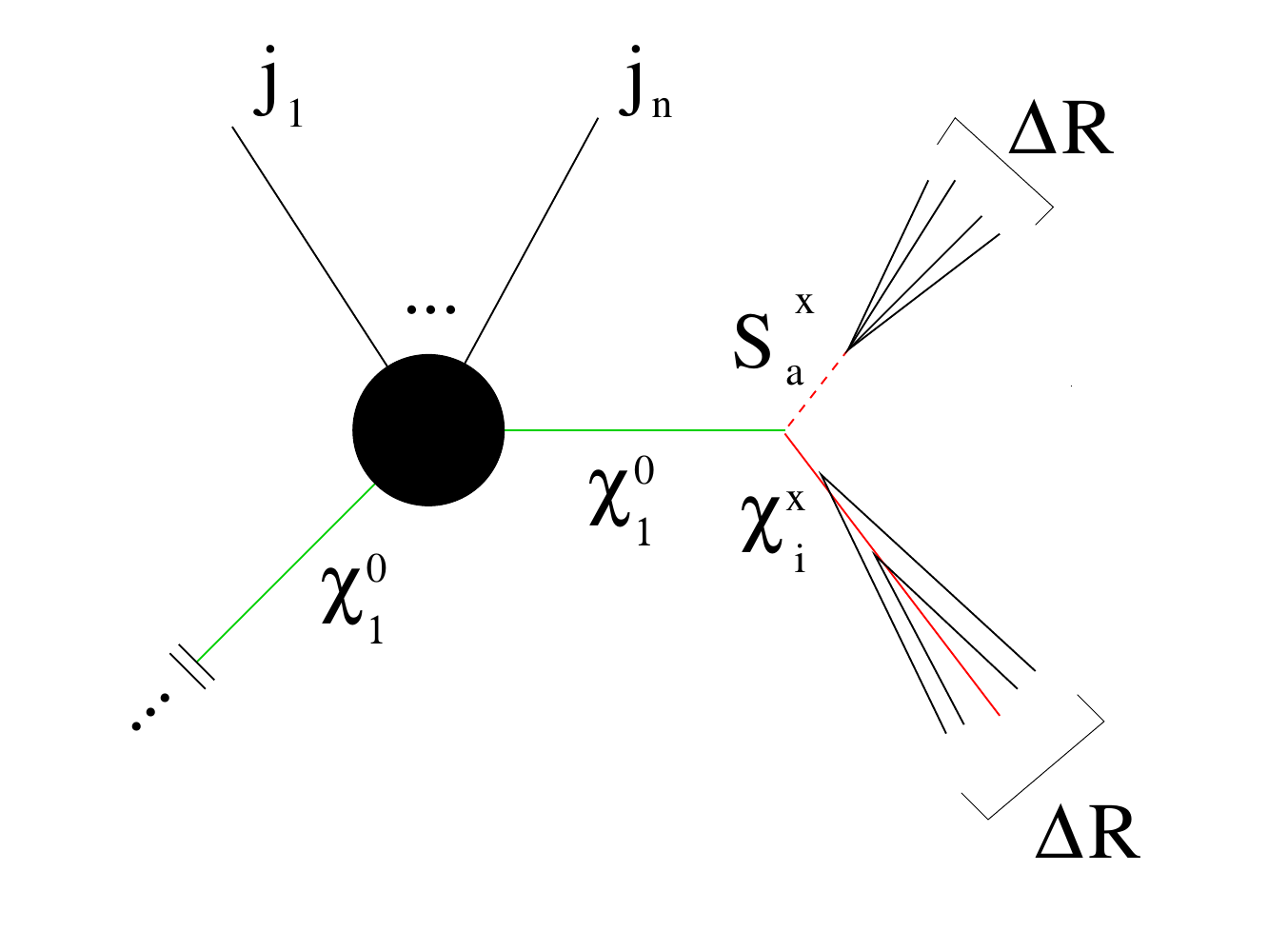}
\end{center}
\caption{Schematic overview of an HV collider event.}
\label{fig:ljet}
\end{figure}

  We can classify supersymmetric events in this scenario
by the number of HV jets they contain together with the structure
of the HV jets themselves.
We illustrate the situation schematically in Fig.~\ref{fig:ljet}.
Each LSMP can produce zero, one, or two HV jets, and there are
two LSMPs per event.
A given HV jet can be prompt, leaving visible tracks directly from
the interaction vertex, or displaced, with visible tracks beginning
only after a macroscopic distance.
An HV jet may also contain the decay products of more
than one vector, yielding additional substructure within the single
collimated detector object.
Thus, we have
\beq
\left\{\begin{array}{c}
\text{MSSM cascade}
\end{array}
\right\}
~\oplus~
\left\{\begin{array}{c}
\text{0 HV jets}\\
\text{1 HV jets}\\
\text{2 HV jets}\\
\text{3 HV jets}\\
\text{4 HV jets}
\end{array}
~\otimes~
\begin{array}{c}
\text{prompt track}\\
\text{delayed track}
\end{array}
\right\}
\eeq
We will show below that many of these possibilities can occur
in different events derived from a single set of hidden sector parameters.

\bigskip

\section{Hidden Sector Parameter Scan\label{sec:scan}}

  As discussed above, the collider signals of a supersymmetric hidden sector
depend on two main quantities: the number HV jets produced per LSMP
decay to the hidden sector; and whether or not there are
any long-lived hidden states that produce displaced vertices.
To investigate the full range of collider signals (for a neutralino LSMP)
and to see how they depend on the hidden sector parameter values, we have implemented this model in FeynRules~1.4.9~\cite{Christensen:2008py} and MadGraph~4.4.49~\cite{Alwall:2007st}.  We have computed decay tables for the LSMP neutralino and hidden sector particles using numerical evaluation of the analytic (tree-level) expressions.  In particular, the decays to the visible sector used the experimentally measured ratio $R$ as described in Eqs.~\eqref{eq:xdecay}, \eqref{eq:xwidthhad} and \eqref{eq:oshelldec}.  For decays to the hidden sector, we have checked our results against BRIDGE~2.23~\cite{Meade:2007js}.


  The parameter space of the hidden sector can be described uniquely
by seven independent parameters (once the spontaneous breakdown
of \Ux\ is imposed).  We have chosen to use the set
$\{\mz,\,\ma,\,M_x,\,\mu',\,\tan\zeta,\,g_x,\,\epsilon\}$.
We then generated $2 \times 10^4$ points in this parameter space,
fixing $g_x=0.3$ and $\epsilon = 10^{-3}$, but otherwise
scanning over the ranges listed in Table~\ref{tab:parscan}.
These values are generally consistent with direct low-energy searches
for light hidden
sectors~\cite{Fayet:2007ua,Pospelov:2008zw,Reece:2009un,Bjorken:2009mm,
Schuster:2009au,Abrahamyan:2011gv,Collaboration:2011zc,Batell:2009di,
Merkel:2011ze,Andreas:2010tp}.
All magnitudes were selected using logarithmic
priors, while the two signs of $\mu'$ were equally probable.
We also took the LSMP to be a mostly-Bino neutralino with a mass of
300\,GeV.  Similar results are expected for mostly Wino or Higgsino
neutralino LSMP provided it has a significant Bino fraction.

\begin{table}[ttt]
  \centering
  \begin{tabular}{|c|c|}
  \hline
  Parameter & Range \\
  \hline
  \mz & (0.1, 10) GeV \\
  \ma & (0.1, 10) GeV \\
  $M_x$ & (0.1, 10) GeV \\
  $\mu'$ & $\pm$ (0.1, 10) GeV \\
  \tz & (0.1, 10) \\
  \hline
  \end{tabular}
  \caption{Range of Hidden Sector parameters considered in the scan.}
\label{tab:parscan}
\end{table}

\subsection{Number of HV Jets}

  The first set of quantities of interest are the probabilities
$\tmp_{\tilde{N}}$ for an LSMP neutralino to produce
$\tilde{N}=0,\,1,\,2$ HV jets.  Additional HV jets can arise
from hidden sector radiation, but we do not consider this effect
here since it is expected to be very mild for the value of $g_x = 0.3$
we are using~\cite{Bai:2009it,Cheung:2009su,Carloni:2010tw}.
Furthermore, for the ranges of hidden sector parameters used in our
scan we find that the $h_1^x$ scalar is almost always effectively stable on collider
scales, either because it is light or due to small mixing angles or
a large boost.  The exceptions comprise $\order{0.1\%}$
of our scan; apart from these points, the only source of HV jets for this range of parameters
is therefore the hidden vector $X^{\mu}$.

  Before addressing the probabilities $\tmp_{\tilde{N}}$
directly, it is instructive to first discuss the two closely-related
quantities $\mathcal{P}_F$ and $\mathcal{P}_B$.  They are defined to be
the probabilities of an HV jet being produced in the hidden sector cascades
initiated by the fermion or the boson decay product of the (neutralino) LSMP,
averaged over all LSMP decay modes and weighted by the relevant branching
fractions.  They are calculated using the hidden and LSMP decay tables
and lengths, specifically by replacing particles by their decay tables
till either an HV jet or a collider-stable particle is produced.
Our working definition of ``collider-stable'' for this study
is a (boosted) decay length greater than $\gamma c\tau > 10\,\text{m}$,
where we estimate the boosts of the hidden states by assuming that they
are produced with an energy of $m_{LSMP}/2 = 150\,\gev$.

\begin{figure}[ttt]
  \begin{center}
    \includegraphics[width=0.5\textwidth]{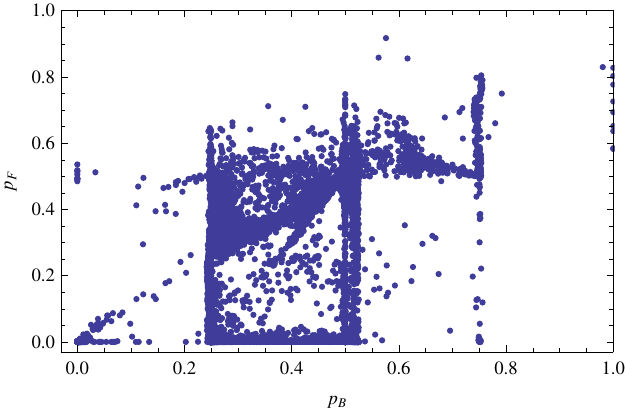}
  \end{center}
  \caption{The distribution of values of $\mathcal{P}_B$ and $\mathcal{P}_F$ found in our parameter scan over the hidden sector.  Each point in the plot corresponds to a single point in the parameter scan, but it is possible for points to overlap at this resolution.}
  \label{fig:pb_vs_pf}
\end{figure}

  In Fig.~\ref{fig:pb_vs_pf} we show the distribution of values
of $\mathcal{P}_B$ and $\mathcal{P}_F$ found in our parameter scan.
Several features are clearly visible in this plot.  In particular, there are
agglomerations of points for: i) $\mathcal{P}_B = 0.25$, 0.5 and 0.75,
with various values of $\mathcal{P}_F$;
ii) $\mathcal{P}_F = 0.0$ and 0.5,
for various values of $\mathcal{P}_B$; and iii) for
$\mathcal{P}_F = \mathcal{P}_B$.
There is also a less-obvious but significant feature at
$\mathcal{P}_B, \mathcal{P}_F \approx 0$ that includes over 40\%
of the points in our scan.  These features can be understood in terms of the
hidden sector spectrum, with mixing effects subdominant.

  The agglomerated values of $\mathcal{P}_B$ are most easily understood in
terms of the result of Eq.~\eqref{eq:lsmpscalar}, that the LSMP decays
to each hidden sector boson with equal frequency.
This implies that the points for which $\mathcal{P}_B = 0$ (0.25, 0.5, 0.75)
are those where none (one, two, three) hidden bosons decay visibly
and the rest invisibly (\emph{i.e.} outside the detector or to
other invisible states).
There are very few points with $\mathcal{P}_B \gtrsim 0.8$ due to $h_1^x$;
as discussed in Sec.~\ref{subsec:decayout}, direct decays of this state
to the SM are normally too slow to occur inside the detector.

\begin{figure}[ttt]
  \begin{center}
    \includegraphics[width=0.5\textwidth]{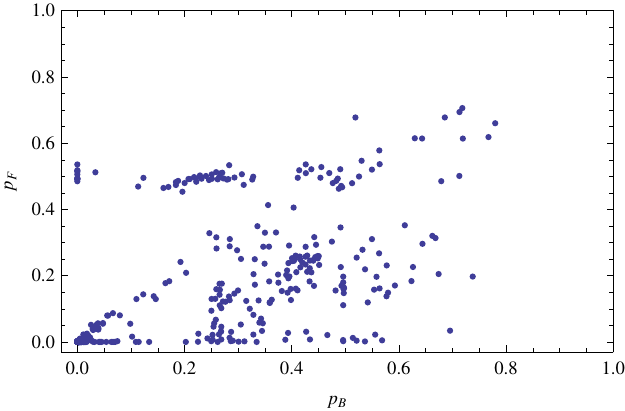}
  \end{center}
  \caption{The distribution of values of $\mathcal{P}_B$ and $\mathcal{P}_F$ from Fig.~\ref{fig:pb_vs_pf}, restricted to those points where the decay $X_\mu \to A^x h^x_1$ is kinematically allowed.  Most points in this plot are at $\mathcal{P}_B, \mathcal{P}_F \approx 0$ and overlap at this resolution.}
  \label{fig:zxtoaxh1}
\end{figure}

  The points in Fig.~\ref{fig:pb_vs_pf} with small values of both
$\mathcal{P}_B$ and $\mathcal{P}_F$ are necessarily those where the
$X^{\mu}$ gauge boson has on-shell decays to other hidden states.
However, the converse is not true.
In Fig.~\ref{fig:zxtoaxh1} we show $\mathcal{P}_B$ and $\mathcal{P}_F$
for those points in our scan where the decay $X_\mu \to A^x h_1^x$ is allowed,
which is almost always the most important hidden mode.
Non-zero values of $\mathcal{P}_B$ still arise in this case when
the pseudoscalar decays sufficiently promptly to an off-shell gauge
boson whose invariant mass is too small to decay invisibly.
When $\mathcal{P}_B$ is small, $\mathcal{P}_F$ will usually also be small
as the fermions can only decay to hidden sector bosons.
The exception occurs when $\chi_2^x$ has no two-body decays,
and will be discussed in more detail below.

  When the $X_{\mu}$ vector boson does decay mainly to the SM,
we find $\mathcal{P}_B\geq 0.25$.  Even larger values of $\mathcal{P}_B$
can come about when the other bosons produce vectors in their decay
chains.  For the pseudoscalar $A^x$, if it has no fermion decay modes
it must decay to either a real or virtual $X_\mu$, which will typically
be visible unless it is far off-shell.
Neglecting subsequent fermion decays, the $A^x$ state thus contributes
\begin{equation}
  \mathcal{P}_B (A^x) \simeq 0.25 \times \biggl[
1 - \sum_{i\geq j} \text{BR} (A_x \to \chi_i^x \chi_j^x)
\biggr] \ .
\end{equation}
In contrast, the heaviest boson $h_2^x$ generically contributes
little to $\mathcal{P}_B$; even if it has no fermionic decays,
the decay to $2h_1^x$ will usually dominate the decays to vectors if
it is allowed.\footnote{Note as well that the decay $h_2^x\to A^x X_\mu$
is never on shell, and if the decay $h_2^x \to 2X_\mu$ is open we are in
the decoupling limit where the scalar decay dominates.}
In particular, the points with $\mathcal{P}_B \gtrsim 0.5$ require
a highly compressed spectrum: $m_{h_1^x} < \ma, \mz < m_{h_2^x} < 2 m_{h_1^x}$,
or in terms of our input parameters $\ma/2 < \mz < 2\ma$ and $0.5 < \tz < 2$.

\begin{figure}[ttt]
  \begin{center}
    \includegraphics[width=0.5\textwidth]{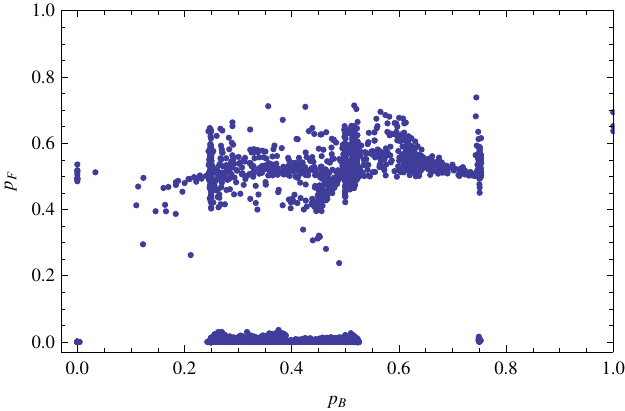}
  \end{center}
  \caption{The distribution of values of $\mathcal{P}_B$ and $\mathcal{P}_F$
in our scan
restricted to those points where the decay
$\chi_2^x \to h^x_1 \chi_1^x$ is \emph{not} kinematically allowed.  The points with $\mathcal{P}_F\approx 0.5$ (0) are those where the $\chi_2^x$ decays inside (outside) the detector.}
  \label{fig:x2offshell}
\end{figure}

  Turning next to the fermion side, the agglomerations of points
at $\mathcal{P}_F \simeq 0,\,0.5$ can arise in several different ways.
First of these is when $\mz <\abs{\mu'} \ll M_x$, leading to a
spectrum with two light, nearly-degenerate Higgsino-like $\chi_1^x,\,\chi_2^x$
states and a more massive gaugino-like $\chi_3^x$.
The LSMP will then decay to $\chi_1^x$ and $\chi_2^x$ with nearly equal
frequency (see Eq.~\eqref{eq:lsmptofermion}) and to $\chi_3^x$ very rarely.
The mass splitting of the Higgsino-like $\chi_{1,2}^x$ pair is roughly
$\mz^2 /\abs{\mu'} < \mz$, so $\chi_2^x$ decays exclusively to either
the invisible $h_1^x$ or an off-shell vector.  The former leads
to $\mathcal{P}_F \simeq 0$; the latter to $\mathcal{P}_F \simeq 0$ or 0.5,
depending on whether the decay occurs inside the detector.
This behaviour can be seen clearly in Fig.~\ref{fig:x2offshell}
where we show the values of $\mathcal{P}_B$ and $\mathcal{P}_F$
for those points where $\chi_2^x\to h_1^x\chi_1^x$ is not kinematically
possible.

\begin{figure}[ttt]
  \begin{center}
    \includegraphics[width=0.5\textwidth]{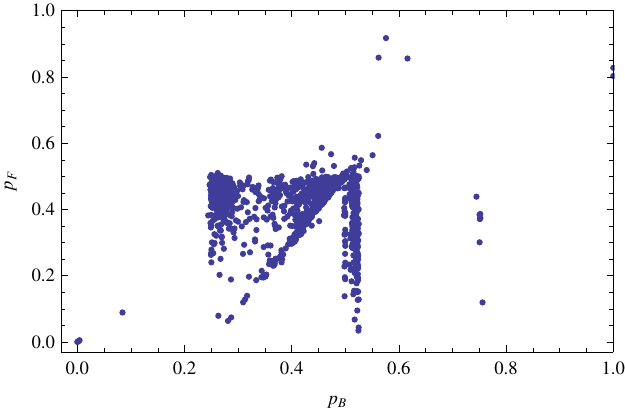}
  \end{center}
  \caption{The distribution of values of $\mathcal{P}_B$ and $\mathcal{P}_F$
in our scan
restricted to those points where the decay
$\chi_2^x \to X_\mu \chi_1^x$ is kinematically allowed and
the decay $\chi_3^x \to A^x \chi_1^x$ is not.}
  \label{fig:x2tozxx1}
\end{figure}

  Values of $\mathcal{P}_F \simeq 0.5$ can also arise when $M_x \ll \abs{\mu'}$ and $\mz \lesssim \abs{\mu'} \lesssim \ma$.  In this case, the Higgsino-like states are the more massive fermions $\chi_2^x$ and $\chi_3^x$, and they decay to the gaugino-like $\chi_1^x$ and either $X_\mu$ or $h_1^x$ exclusively.  While the branching ratios for one Higgsino to these two final states may be unequal, generically the other Higgsino has a complementary decay pattern.  The result is that the rates for $\chi_1^0 \to \chi_{2,3}^x \to S^x \chi_1^x$ are equal for $S^x = X_\mu, h_1^x$,
leading to $\mathcal{P}_F \simeq 0.5$.  This feature is illustrated
in Fig.~\ref{fig:x2tozxx1}, where we show the values of $\mathcal{P}_B$ and
$\mathcal{P}_F$ for those points where $\chi_2^x \to X_\mu \chi_1^x$ is
kinematically allowed but $\chi_3^x \to A^x \chi_1^x$ is not.

\begin{figure}[ttt]
  \begin{center}
    \includegraphics[width=0.5\textwidth]{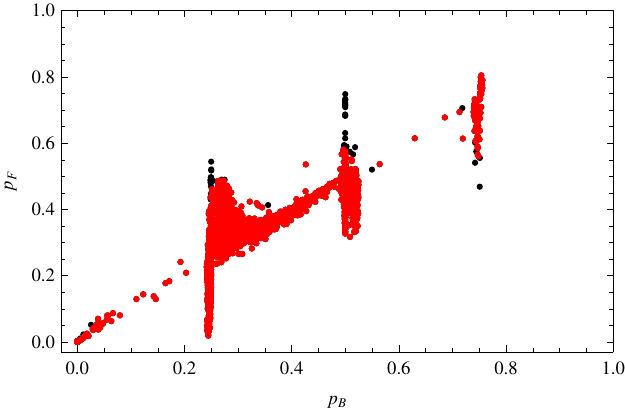}
  \end{center}
  \caption{The distribution of values of $\mathcal{P}_B$ and $\mathcal{P}_F$
obtained in our scan
restricted to those points where the decays
$\chi_2^x \to X_\mu \chi_1^x$ and $\chi_2^x \to A^x \chi_1^x$
are kinematically allowed.
The points in black (red) are those where the decay
$\chi_2^x \to h_2^x \chi_1^x$ is (is not) allowed.}
  \label{fig:equal}
\end{figure}

  In Fig.~\ref{fig:pb_vs_pf} we also see an agglomeration of points with
$\mathcal{P}_F \sim \mathcal{P}_B$.
This typically occurs when the Higgsino-like fermions are somewhat heavy
and can decay to all scalars.
The fermionic side of the LSMP decay chain then proceeds according
to $\chi_1^0 \to \chi_{2,3}^xS^x$, followed by $\chi_{2,3}^x\to \chi_1^x{S^x}'$
with roughly equal probability for all four bosonic final states ${S^x}'$.
Thus, the distribution of bosons produced in the decays of $\chi_{2,3}^x$
in this case mirrors those of the neutralino LSMP giving
$\mathcal{P}_B\sim \mathcal{P}_F$.  We illustrate this behaviour in
Fig.~\ref{fig:equal}, where we show in black the values of
$\mathcal{P}_B$ and $\mathcal{P}_F$ for those points where
$\chi_2^x\to \chi_1^x (X_{\mu}$, $h_1^x$, $A^x$, $h_2^x$) are all allowed.
The additional agglomerations of points in Fig.~\ref{fig:equal}
extending away from the $\mathcal{P}_F \simeq \mathcal{P}_B$ line
correspond to parameter values that lead to all fermions being
heavier than the bosons but also well-mixed.
In Fig.~\ref{fig:equal} we also show in red those points
where $\chi_2^x\to \chi_1^xX^{\mu}$ and $\chi_2^x\to\chi_1^x A^x$
are allowed but $\chi_2^x\to \chi_1^x\,h_2^x$ is not.
These points have nearly the same structure as those
where all three boson modes are open since, as discussed above,
the decays of $h_2^x$ are typically invisible.

\begin{figure}[htt]
  \begin{center}
    \includegraphics[width=0.5\textwidth]{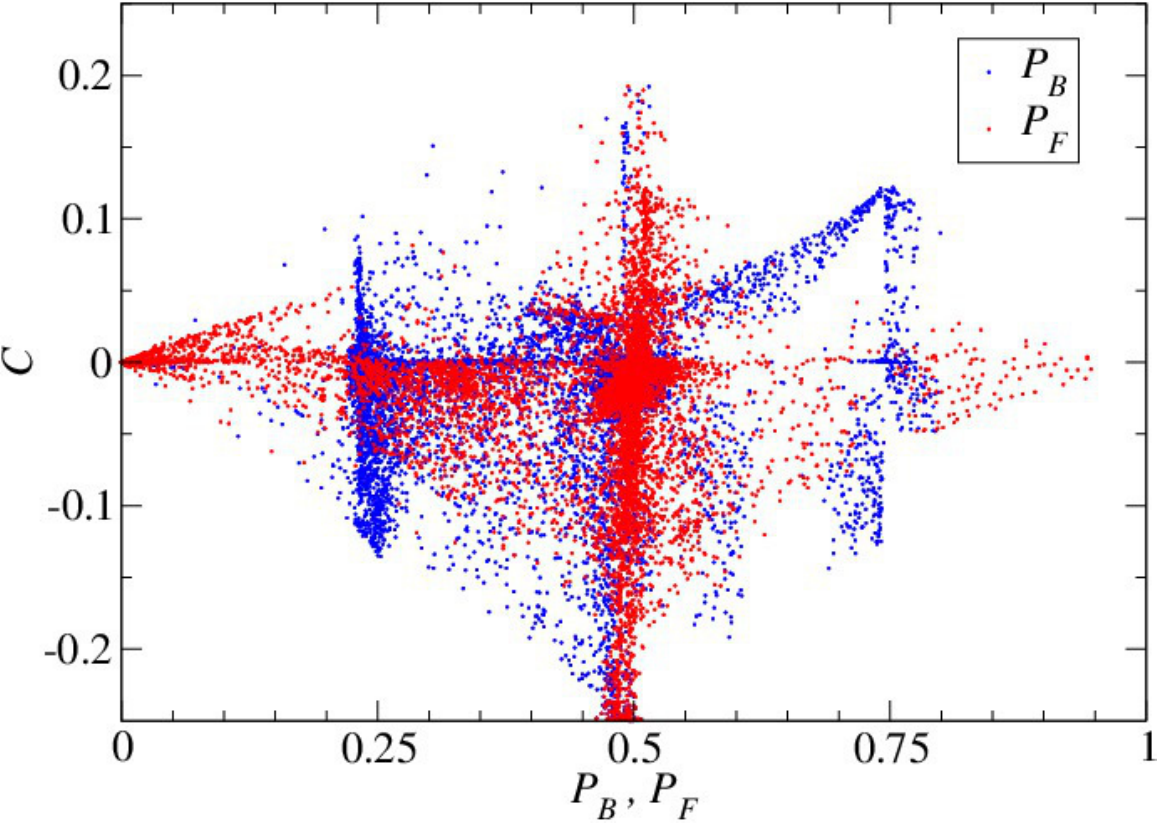}
  \end{center}
  \caption{Values of $\tmp_B$ and $\tmp_F$ relative to the
correlation coefficient $\mathcal{C}$.}
  \label{fig:pbpfcorr}
\end{figure}

  The probabilities $\mathcal{P}_{B,F}$ are related to the
probabilities $\tmp_{\tilde{N}}$ that a LSMP neutralino
decay will produce $\tilde{N} = 0$\,--\,2 HV jets:
\bea
\tmp_0 &=& 1-\mathcal{P}_B-\mathcal{P}_F
+(\mathcal{P}_B\mathcal{P}_F-\mathcal{C})\ ,
\label{eq:tmp0}\\
\tmp_1 &=& \mathcal{P}_B+\mathcal{P}_F
-2(\mathcal{P}_B\mathcal{P}_F-\mathcal{C})\ ,
\label{eq:tmp1}\\
\tmp_2 &=& \mathcal{P}_B\mathcal{P}_F-\mathcal{C} \ ,
\label{eq:tmp2}
\eea
where $\mathcal{C}$ describes the degree of correlation between
$\mathcal{P}_B$ and $\mathcal{P}_F$ (with Eq.~\eqref{eq:tmp2}
as its defining relation).  In particular, if $\mathcal{P}_B$
and $\mathcal{P}_F$ were independent, $\mathcal{C}$ would vanish.
 In Fig.~\ref{fig:pbpfcorr} we show the values of the correlation coefficient
$\mathcal{C}$ relative to $\mathcal{P}_B$ and $\mathcal{P}_F$.

  By $R$-parity, each supersymmetric event will involve two neutralino
LSMPs.  Since the two LSMP decays in an event
are independent, it is trivial to relate the values of
$\tmp_{\tilde{N}}$ to the probabilities $P_{N}$ for an event
to have $N=0\!-\!4$ HV jets.\footnote{To wit:
~~$P_0 = \tmp_0^2,~~~P_1=2\tmp_0\tmp_1,~~~P_2=\tmp_1^2+2\tmp_0\tmp_2,~~~
P_3=2\tmp_1\tmp_2,~~~P_4=\tmp_2^2.$}
Based on our previous results, we can identify three qualitative behaviours:
\begin{itemize}
  \item Small $\mathcal{P}_B$ and $\mathcal{P}_F$, with the number of HV jets peaked
at $N=0\!-\!1$;
  \item Large $\mathcal{P}_B$ and $\mathcal{P}_F$, with the number of HV jets peaked at
$N= 3\!-\!4$;
  \item Intermediate cases, with the number of HV jets peaked around $N=2$.
\end{itemize}
In the last case, there is a further distinction between
$\mathcal{P}_B \sim \mathcal{P}_F$,
where the distribution in the number of HV jets is frequently broad;
and $\mathcal{P}_B \gg \mathcal{P}_F$ or $\mathcal{P}_B \ll \mathcal{P}_F$,
when the distribution is narrow.
Considering the classification by decay mode discussed above,
we note that approximately 40\%
of points with $\mathcal{P}_F, \mathcal{P}_B \simeq 0$
fall into the first category; most of the points with
$\mathcal{P}_B \simeq 0.75$ fall into the second category;
and the remaining points along the $\mathcal{P}_F \simeq \mathcal{P}_B$
line fall into the last category.  The points with invisible fermions
but moderate to large $\mathcal{P}_B$ also fall into the last category,
but into the ``narrow'' subset.

\subsection{Presence of Displaced Vertices}

The other main phenomenological feature of the hidden sector
is the presence or absence of displaced vertices.
Our working definition of a displaced vertex is a decay from the
hidden sector to the visible with a (boosted) decay length larger
than $\gamma\,c\tau > 0.1\,\text{mm}$
but less than $\gamma\,c\tau = 10\,\text{m}$.
This definition is crude, as the phenomenological implications depend
in a very important way on the location in the detector the decay occurs.
For example, it is much easier to resolve a displaced vertex if the products
are visible in the tracking chambers (corresponding to a true displaced
vertex in the usual sense)~\cite{Ball:2007zza,Aad:2009wy}.
We will mostly ignore this distinction
in this section, but we shall discuss it in more detail
in the next.

\begin{figure}[ttt]
  \begin{center}
    \includegraphics[width=0.47\textwidth]{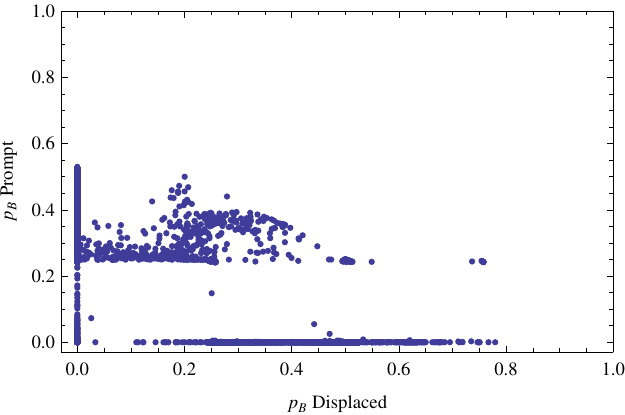}
\hspace{0.5cm}
\includegraphics[width=0.47\textwidth]{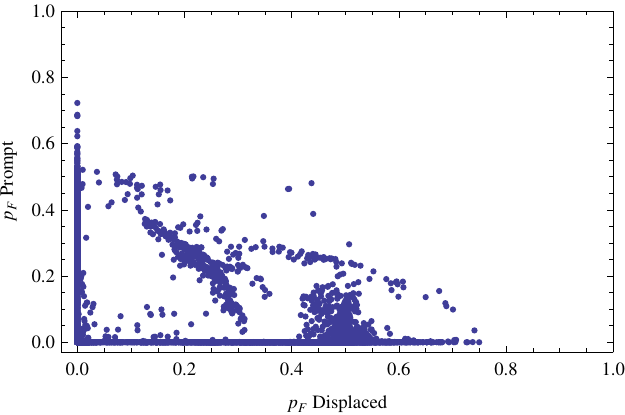}
  \end{center}
  \caption{Relative frequencies of prompt and displaced decays
for the bosonic (left) and fermionic (right) branches of
LSMP neutralino decay cascades.}
  \label{fig:disp}
\end{figure}

  The relative probabilities for prompt and displaced vertices from the
bosonic (fermionic) side of the decay chain are shown in the left (right)
panel of Fig.~\ref{fig:disp}.
In particular, note that the sum of the two probabilities plotted equal
$\mathcal{P}_B$ ($\mathcal{P}_F$).
Both plots show concentrations of points along the axes where all decays
are either prompt or displaced, as well as a numerically smaller set
involving both types of decays.

  For the moderate value of $\epsilon=10^{-3}$ we are using,
we find that displaced vertices correspond to off-shell decays
in the hidden sector.  More precisely displaced vertices only occur
if either $A^x$ or $\chi_2^x$ have no two-body decays.
We also find some points in our scan where $h_2^x$ also decays off-shell,
but the structure of the scalar potential means that for all such
points the same is true for $A^x$.  There are no points in our scan where
$\chi_3^x$ has no two-body decays.

The points where all decays are displaced are the simplest to understand;
they are almost exclusively the points (shown in Fig.~\ref{fig:zxtoaxh1})
where the hidden vector can decay into the hidden sector.
This is reasonable as in such a scenario the \emph{only} significant
decays from the hidden sector to the SM must be those where $X_\mu$
is off-shell.  There are some additional points where the fermionic
decays are all displaced that correspond to a subset of the points
plotted in Fig.~\ref{fig:x2offshell}, where the only fermionic
decays back to the SM are $\chi_2^x \to X_\mu^\ast \chi_1^x$.
The set of points where all decays are prompt is also qualitatively simple,
but involves a large number of different possible spectra and decay patterns.
In essence, however, they correspond to points where all hidden sector
particles either decay on-shell or are collider stable.

  More interesting are the points with both prompt and displaced
vertices.
On the bosonic side, there are two obvious features in the left panel of
Fig.~\ref{fig:disp}.  The line of points with
$\mathcal{P}_B\text{(prompt)}\simeq 0.25$
are those where the hidden gauge boson decays promptly to the SM,
while $A^x$ and $h_2^x$ have displaced off-shell decays,
on-shell invisible decays,
or decays to $\chi_2^x$ which in turn decays off-shell.
Note that even when these heavier scalars can decay to the hidden vector,
such modes are suppressed by mixing angles in the decoupling limit
so that decays to fermions will typically dominate when they are open.
The other main feature on the bosonic side are the points with
$\mathcal{P}_B\text{(prompt)}>0.25$.   These also involve decays
to a long-lived $\chi_2^x$ state, but now the heaviest fermion
$\chi_3^x$ is also produced in scalar decays.
It decays promptly, emitting vectors some fraction of the time.
This last feature is more sensitive to the fermion mixing matrices,
which is why the corresponding points form an amorphous blob
instead of a line.

  The fermionic decays that lead to both prompt and displaced
vertices have a more complex structure, as can be seen in the right
panel of Fig.~\ref{fig:disp}.   This structure decomposes into three parts.
The first is a collection of points at
$\mathcal{P}_F\text{(displaced)}\simeq 0.5$ that arise from
off-shell $\chi_2^x$ and on-shell $\chi_3^x$ decays,
though the latter are suppressed by small LSMP branching fractions.
The second fermionic feature is the line of points where the
prompt and displaced probabilities sum to about 0.5.
This subset is characterised by off-shell $A^x$ and on-shell
$h_2^x$ decays, plus heavier Higgsino-like fermions able to decay to
all the bosons on-shell.
The third feature in Fig.~\ref{fig:disp} are the two lines sloping
between $\tmp_F(\text{prompt})$ and $\tmp_F(\text{displaced})$
from $(0.0,0.5)$ to $(0.2, 0.4)$ and $(0.2, 0.8)$.
These points have similar spectra to those of the second feature,
except that now the heaviest scalar $h_2^x$ is also constrained to
decay through an off-shell gauge boson.

\bigskip

\section{Sample Points\label{sec:bench}}

  To further illustrate the diverse range of collider signals that can
arise from this minimal HV theory, we present and investigate the
properties of five distinct sample parameter points obtained in our scan.
As in Section~\ref{sec:scan}, we assumed a Bino-like neutralino
LSMP of mass $m_{\chi_1^0} = 300\,\gev$.

\subsection{General Properties\label{subsec:benchP04}}

  In Table~\ref{tab:bench} we list the values of
$\mathcal{P}_B$, $\mathcal{P}_F$,
$\tmp_0$, $\tmp_1$, and $\tmp_2$ for all five sample points.
We also list the characteristic decay lengths $c\tau$ of
any long-lived particles likely to lead to displaced vertices.
Note that since the LSMP is much heavier than the hidden sector states,
the observed displacements will be greater by a factor
$\gamma \sim m_{LSMP}/2m_{hid} \sim \order{50\,\text{--}\,100}$.
In Table~\ref{tab:spectrum} we list
the mass spectrum for the five benchmark points, and the
underlying parameter input values are tabulated in Appendix~\ref{appb}.

\begin{table}[ttt]
  \centering
  \begin{tabular}{|c|c|c|c|c|c|c|}
    \hline
    & $\mathcal{P}_B$ & $\mathcal{P}_F$ & $\tmp_0$ & $\tmp_1$ & $\tmp_2$ & $c\tau$(mm) \\
    \hline
    HV1 & 0.00 & 0.00 & 1.00 & 0.00 & 0.00 & 0.0 \\
    HV2 & 0.25 & 0.00 & 0.75 & 0.25 & 0.00 & 0.0 \\
    HV3 & 0.26 & 0.28 & 0.57 & 0.31 & 0.12 & 10.0 \\
    HV4 & 0.46 & 0.44 & 0.34 & 0.43 & 0.23 & 0.0 \\
    HV5 & 0.50 & 0.57 & 0.22 & 0.50 & 0.28 & 0.0, 0.2, 8.4  \\
    \hline
  \end{tabular}
  \caption{Decay yields $\tmp_B$ and $\tmp_F$ and decay probabilities
$\tmp_0$, $\tmp_1$, $\tmp_2$ from LSMP neutralino decays for all five
sample points as well as the 
decay lengths $c\tau$ of any particles likely to lead to displaced vertices
in high-energy collider events.}
\label{tab:bench}
\end{table}

\begin{table}[ttt]
  \centering
  \begin{tabular}{|c|c|c|c|c|c|c|c|}
    \hline
    & $m_x$ & $m_{h_1^x}$ & $m_{h_2^x}$ & $m_{A^x}$ & $m_{\chi_1^x}$ & $m_{\chi_2^x}$ & $m_{\chi_3^x}$ \\
    \hline
    HV1 & 5.73 & 0.57 & 5.73 & 0.60 & 1.91 & 9.28 & 9.50 \\
    HV2 & 0.65 & 0.06 & 7.46 & 7.43 & 0.57 & 1.09 & 1.34 \\
    HV3 & 6.49 & 1.95 & 7.40 & 4.04 & 2.07 & 6.42 & 7.28 \\
    HV4 & 0.90 & 0.67 & 9.74 & 9.72 & 0.48 & 1.93 & 2.12 \\
    HV5 & 3.54 & 1.82 & 4.26 & 3.00 & 1.97 & 3.70 & 10.14 \\
    \hline
  \end{tabular}
  \caption{Particle masses in GeV for the sample points
discussed in the text.}
  \label{tab:spectrum}
\end{table}

\subsubsection{HV1 -- Invisible}
  The first benchmark point HV1 is essentially invisible, with nearly
all hidden sector cascades producing only states that are stable
($\chi_1^x$) or very long-lived on collider timescales
($h_1^x$, $A^x$
).  The vector decays almost entirely invisibly via
$X^{\mu}\to h_1^xA^x$ rather than to the SM ($BR \lesssim 10^{-6}$).
The lighter two of the other bosonic states decay slowly through
multi-body modes and are effectively stable on collider time scales,
and the lightest fermion is stable and also invisible.
The heaviest boson and the two heavier fermions decay to
the invisible states, not the vector.
Thus, the hidden states produced by the cascades just contribute
to missing energy, and this theory is nearly indistinguishable from
the MSSM at high-energy colliders.

\subsubsection{HV2 -- Visible Vectors}
  This point yields hidden cascades that are invisible except
for when the vector $X^{\mu}$ arises directly in the LSMP decay.
Once produced, the vector decays promptly and nearly entirely to the SM\@.
The other particles are stable or long-lived, or decay to invisible states.

  About one quarter of the LSMP neutralino decays produce a vector,
yielding a probability distribution
for the total number of HV jets in supersymmetric events of
\beq
P_0 = 0.56,~~P_1 =0.38 ,~~P_2 = 0.06,~~P_3 = 0,~~P_4 = 0 \ .
\nnmb
\eeq
We see that over half the supersymmetric events for this point
reproduce the standard MSSM pattern of missing energy, and therefore
the standard supersymmetry searches will be sensitive to the MSSM
portion of this HV theory as well~\cite{Ball:2007zza,Aad:2009wy}.
However, over a third of the supersymmetric events will also contain
an additional HV jet beyond the standard MSSM cascade products.
Detecting and measuring such an object would give evidence for
the existence of an additional hidden sector.

\subsubsection{HV3 -- Displaced Vertex}
  The third sample point has a more complicated decay pattern as well
as a macroscopically displaced decay vertex.
Both the boson and fermion sides of the LSMP decay chain can
produce HV jets.  On the bosonic side, visible HV jets come mainly
from direct decays to the pseudoscalar followed by
$A^x\to h_1^x{X^{\mu}}^*$ with the off-shell vector going to the SM\@.
This decay has a macroscopic lifetime of $c\tau = 10\,\text{mm}$ due to
the vector being far off shell, but the actual decay length in collider events
will be enhanced by a relativistic boost of
$\gamma \sim m_{\chi_1^0}/m_{A^x} \sim \order{50}$.
Similarly, HV jets are produced on the fermionic side from the $A^x$
produced in the decays $\chi_{2,3}^x\to \chi_1^xA^x$.

  In total, the distribution of HV jets in supersymmetric events is
\beq
P_0 = 0.32,~~P_1 = 0.36,~~P_2=0.23,~~P_3=0.07,~~P_4=0.01 \ .
\nnmb
\eeq
Again, a significant fraction of these events produce missing
energy as in the MSSM, but the majority also have HV jets characterised
by a displaced vertex.

\subsubsection{HV4 -- Multiple HV Jets}
This sample point is very likely to produce additional HV jets.
The vector $X^{\mu}$ decays promptly to the SM nearly all the time.
In addition to direct production from the LSMP, vectors are frequently
produced in the decays of the two Higgsino-like heavier fermions.
The heavier scalars rarely decay directly to $X^{\mu}$;
instead they dominantly decay to fermions,
producing HV jets through a multi-step decay chain.

  The probability distribution for HV jets in supersymmetric events is
\beq
P_0 = 0.11,~~P_1 = 0.29,~~P_2=0.34,~~P_3=0.20,~~P_4=0.05 \ .
\nnmb
\eeq
In this case, only a small fraction of the collider events
fall into the usual MSSM form.

\subsubsection{HV5 -- Multiple HV Jets and Displaced Vertices}
  Our fifth and last benchmark point produces many HV jets
with both prompt and displaced decay vertices.
The vector decays promptly to the SM almost exclusively.
The pseudoscalar decays mainly through $A^x\to h_1^x{X^{\mu}}^*$
while the heavy scalar decays mostly invisibly via $h_2^x\to \chi_1^x \chi_1^x$.
The pseudoscalar decay length is $c\tau = 8.4$~mm.
On the fermionic branch of the LSMP decay, the two lighter Higgsino-like
states are produced most abundantly.  Of these, the decay of the
heavier state $\chi_2^x\to \chi_1^x{X^{\mu}}^*$ involves an
off-shell vector and has a decay length of $c\tau = 0.2$~mm.
Again, both the metastable states will be highly boosted when produced
in LSMP decays, increasing the observed displacements.

  The distribution of the number HV jets produced by MSSM collider events is
\beq
P_0 = 0.05,~~P_1 = 0.22,~~P_2=0.37,~~P_3= 0.28,~~P_4 =0.08  \ .
\nnmb
\eeq
Only a small fraction of collider events will have no additional
HV activity, while those events with HV jets can have both prompt and
displaced decay vertices.

\subsection{Structure of HV Jets}

  The basic structure of the HV jets arising for all the sample points
is similar and relatively simple.  In most cases, they come from the decay
of a \emph{single} hidden vector $X^{\mu}$ boosted by an amount
$\gamma \simeq m_{\chi_1^0}/2{m_1}$, where $m_1$ is the mass of first
boson or fermion produced in the LSMP neutralino decay chain that gives
rise to the vector.  The exception is HV4, where the two heavier scalars
can decay to two vectors through intermediate fermions; however, even
here the single vector channels are more common.
As such, each HV jet is typically initiated by two
hard objects collimated within a cone of size
$\Delta R \lesssim \gamma\,(\sqrt{s_x}/m_1)$, where
$\sqrt{s_x}\leq m_x$ is the invariant mass of the (possibly off-shell)
decaying vector.
More complicated HV jets containing the decay products of multiple vectors
could arise if we took a larger coupling $g_x$ to produce significant
hidden vector showering.

  The content of the HV jets depends on the value of $\sqrt{s_x}$
according to Eq.~\eqref{eq:xdecay}.  We have
$\sqrt{s_x} = m_x$ for on-shell decays (HV2, HV4, HV5);
$\sqrt{s_x} \leq (m_{A^x}-m_{h_1^x})$ for off-shell pseudoscalar decays
(HV3 and HV5); and $\sqrt{s_x} \leq (m_{\chi_2^x}-m_{\chi_1^x})$
for off-shell fermion decays (HV5).

  Leptonic HV jets are the most promising collider signature of this
class of supersymmetric hidden valleys.  Their identification will
likely require a dedicated search since the leptonic ($e,\,\mu$)
constituents are expected to be nearly collinear and fall within
a single typical lepton isolation cone ($\Delta R \leq 0.4$).
Proposals for such searches have been made for the Tevatron~\cite{Abazov:2009hn,Abazov:2010uc,Baumgart:2009tn,Cheung:2009su,Falkowski:2010gv}
and the LHC~\cite{Baumgart:2009tn,Cheung:2009su,Falkowski:2010cm}.
For example, Ref.~\cite{Cheung:2009su} advocated
a requirement of two or more leptons with $p_T > 10\,\gev$
collimated within a cone of $\Delta R < 0.1$ with an additional
isolation cut of $\sum p_T < 3\,\gev$ in the surrounding region
$0.1<\Delta R < 0.4$.

  This type of search is expected to be highly efficient
for the prompt leptonic HV jets arising for HV2, HV4, and HV5,
especially because they derive from a single hidden vector decay and have
a relatively simple substructure, additional radiation
or substructure tends to degrade the efficiency of these
cuts~\cite{Cheung:2009su,Falkowski:2010cm}.
It would also be interesting to study the prospects of identifying
tauonic or hadronic HV jets that also arise frequently for these sample
points based on their collimated substructure, but such an investigation
is beyond the scope of the present work.

  Benchmark points HV3 and HV5 contain displaced vector decays.
These originate from $A^x\to h_1^x{X^{\mu}}^*$ with the vector off shell
for both points, and from $\chi^x_{2}\to \chi_1^x{X^{\mu}}^*$
for HV5.  Taking into account the net boost from the initial LSMP
neutralino decay, these decays are displaced from the collision vertex
by $\gamma c\tau \sim 10\,\text{--}\,50$~cm for HV3 and
$\gamma c\tau \sim 10\,\text{--}\,50$,
0.1\,--\,1~cm ($A^x$ and $\chi_{2}^x$) for HV5.
Such decay lengths will lead to displaced vertices in the tracking
systems of the ATLAS and CMS detectors\cite{Ball:2007zza,Aad:2009wy}.
Even so, the small number of associated tracks (typically two) and
the low invariant masses of the vector decay products could make
these displaced decays difficult to distinguish from photon conversions
in the detectors~\cite{Ball:2007zza,Aad:2009wy,Aad:2011zb}.
Longer decay lengths, emerging outside the ATLAS and CMS tracking
chambers, would be much more difficult to identify.

\subsection{Collider Signals\label{sec:coll}}

  Our analysis thus far has been based on the particle decay tables for
the hidden sector and LSMP.  While a full simulation of experimental
factors is beyond to scope of the current paper, we have initiated
a parton-level analysis of events at the 7~TeV LHC.
Since our focus is on the hidden valley phenomenology, we used a
simple MSSM spectrum with a 300~GeV Bino-like neutralino, an 800~GeV gluino,
and all other new states heavy ($\gtrsim 2.5$~TeV).
This is consistent with current LHC bounds from conventional supersymmetry
searches~\cite{ATLASPrePrint,CMSPrePrint}, but will be probed by upcoming
searches in the near future.  The dominant production channel for these
parameters is $pp \to \tilde{g}\tilde{g}$, with a leading-order
production cross section of about $200$~fb.  All other supersymmetric
production processes are negligible, and in particular the pair production
of the Binos has a cross section of $0.005$~fb.
Once created, the gluinos will decay to $q\bar{q}\chi_1^0$,
so in addition to the two LSMP decay products discussed
above, there will typically be four hard QCD jets in the final state.

  For each of our sample points, we generated 50\,000 events using
MadGraph~4.4.49~\cite{Alwall:2007st}, and then used
BRIDGE~2.23~\cite{Meade:2007js} to decay the gluino, Bino, and hidden
sector particles.  We modified the BRIDGE decay tables by hand to correctly include the hadronic contributions to the width.  To simulate detector acceptances and efficiencies,
we imposed a variety of parton-level cuts.  All objects with pseudo-rapidity
$\abs{\eta} > 2.5$ were ignored.  We defined an HV jet as consisting
of at least two hard subobjects each with $p_T > 10$~GeV collimated
within a cone of size $\Delta R < 0.1$ with an isolation cut of
$\sum p_T < 3$~GeV within $0.1<\Delta R < 0.4$ relative to the net momentum
of the two hardest subobjects in the central cone.
However, we did not distinguish between leptonic and hadronic HV jets,
nor did we account for the some of the hidden decay products being displaced.
Thus, our estimates of HV jet acceptance are likely to be overly optimistic.
We also defined a (parton-level) QCD jet as consisting of one or more
objects within $\Delta R = 0.4$ that fail the HV jet criteria and have a total
$p_T > 20$~GeV.  Lastly, we define the missing transverse energy
\met\ as the sum of the transverse momenta of all jets and HV jets.

\begin{figure}[tt]
  \begin{center}
    \includegraphics[width=0.5\textwidth]{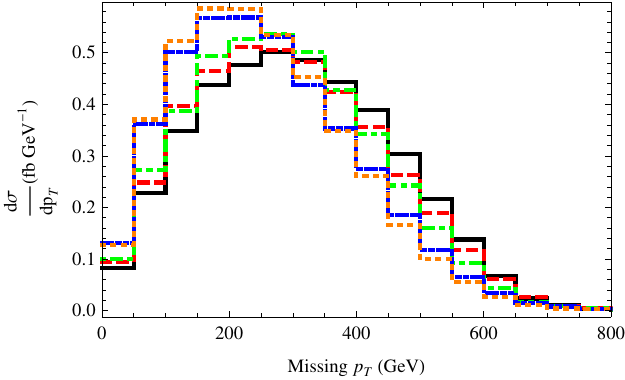}
  \end{center}
  \caption{The \met\ distributions for the five benchmark points
for gluino-initiated events.  The solid black (red dashed, green dot-dashed, blue dot-dot-dashed, orange dotted) line corresponds to HV1 (HV2, HV3, HV4, HV5).}
  \label{fig:misspt}
\end{figure}

  In Fig.~\ref{fig:misspt} we show the distributions of missing energy
for gluino-initiated events in sample points HV2--HV5.  This includes the contribution from metastable hidden states decaying outside the detector.  As already noted,
the presence of several hard jets from the gluino decays will make such
events easy to trigger on even when the hidden sector is completely invisible.
We see that as the expected number of HV jets increases from HV1 to HV5,
the missing energy distribution peaks at lower values.
However, even for the most active benchmark points, the majority of
supersymmetric events still have a significant amount of missing energy
implying that the standard search channels for supersymmetry at the
LHC will be sensitive to these sample points as well.

We show the tagging efficiencies for HV jets for HV2--5 in Table~\ref{tab:HVeff}
(recall that HV1 does not produce HV jets).
Some HV jets will fail to tag due to overlapping with other HV jets
or the hard jets from gluino decay.  However, the difference
in tagging efficiency for the different benchmarks is mostly due
to differences in the LSMP decay chains.  Note that the tagging
efficiency is much higher for HV2, where all HV jets come from
the direct production of the hidden vector from the Bino.
In contrast, the other three benchmarks involve longer decay chains
through the hidden sector (HV4), three body decays out of
the hidden sector (HV3) or both (HV5).
Both effects lead to softer final state particles,
and the products of three-body decays tend to be less collinear
and have a broader spread in energy than when this decay is from
an on-shell vector.  This increases the fraction of candidate HV jets
that are either too separated, $\Delta R > 0.1$, or fail to have
two sufficiently hard ($p_T>10\,\gev$) seed partons.

\begin{table}[ttt]
  \centering
  \begin{tabular}{|c|c|}
    \hline
    Benchmark & HV Jet Tag \\
     & Efficiency \\
    \hline
    HV2 & 88\% \\
    HV3 & 50\% \\
    HV4 & 63\% \\
    HV5 & 54\% \\
    \hline
  \end{tabular}
  \caption{Efficiencies to tag HV jets for the four benchmark points
that produce them according to the requirements listed in the text.}
  \label{tab:HVeff}
\end{table}

\begin{figure}[ttt]
  \begin{center}
    \includegraphics[width=0.47\textwidth]{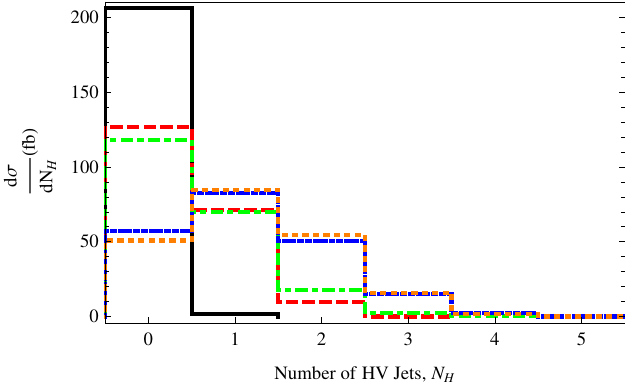}
    \hspace{0.5cm}
    \includegraphics[width=0.47\textwidth]{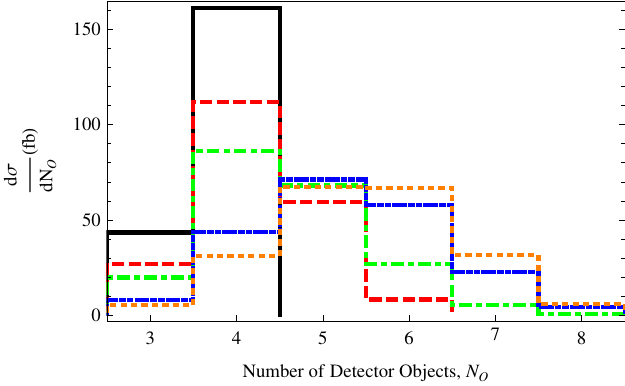}
  \end{center}
  \caption{The distribution of the number of HV jets (left) and HV jets plus jets (right) in gluino-initiated events for our benchmark points.  The solid black (red dashed, green dot-dashed, blue dot-dot-dashed, orange dotted) lines correspond to HV1 (HV2, HV3, HV4, HV5).}
  \label{fig:numobj}
\end{figure}

We next show the distributions of the number of HV jets,
and of distinct (parton-level) jets and HV jets combined
in Fig.~\ref{fig:numobj}.
Compared to the values of $P_0 \ldots P_4$ computed in
Sec.~\ref{subsec:benchP04}, the HV jet number distribution here
is smaller as expected from Table~\ref{tab:HVeff}.  However,
by examining the distributions including (parton-level) jets
we can see that a sizeable fraction of would-be HV jets are
still tagged as jets.  This suggests that the tagging efficiency
could be improved based on a closer study of the substructure of these objects.

\begin{figure}[tt]
  \begin{center}
    \includegraphics[width=0.47\textwidth]{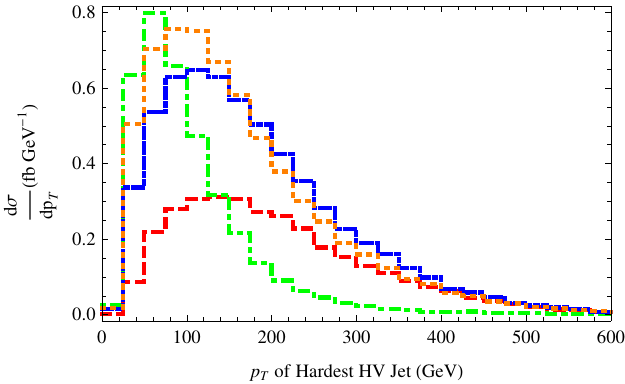}
    \hspace{0.5cm}
    \includegraphics[width=0.47\textwidth]{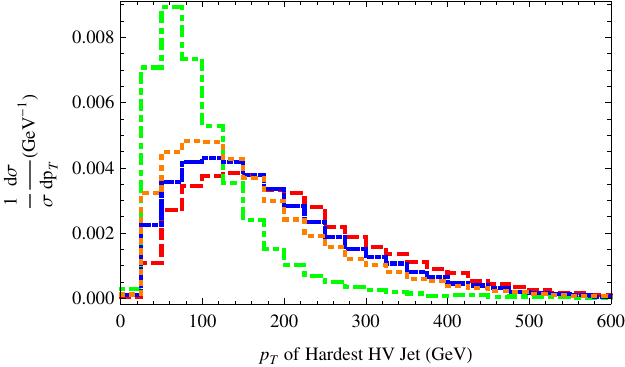}
  \end{center}
  \caption{The $p_T$ distributions for the hardest HV jets in gluino-initiated events.  Left: the distributions for absolute cross-sections.  Right: the distributions normalised to the total cross-section for events with at least one HV jet.  The red dashed (green dot-dashed, blue dot-dot-dashed, orange dotted) lines correspond to HV2 (HV3, HV4, HV5).}
  \label{fig:hardHVpt}
\end{figure}

In Fig.~\ref{fig:hardHVpt} we show the $p_T$ distributions of
the hardest HV jet in a given event for HV2--5.  When we normalise
the distribution to the total cross section for events with at least
one HV jet, it is obvious that HV3 produces much softer HV jets
than HV2, HV4 and HV5, with HV2 producing the hardest objects.
This is caused by the same effects that lower the HV
jet tagging efficiency, \emph{i.e.} higher multiplicity of the final state.
In Appendix~\ref{appc} we show the $p_T$ distributions for all
HV jets produced in these benchmarks.

\begin{figure}[htt]
  \begin{center}
    \subfigure[HV3]{\includegraphics[width=0.47\textwidth]{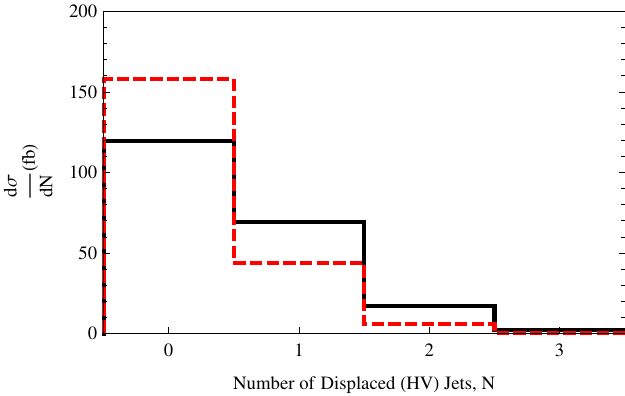}}
    \subfigure[HV5]{\includegraphics[width=0.47\textwidth]{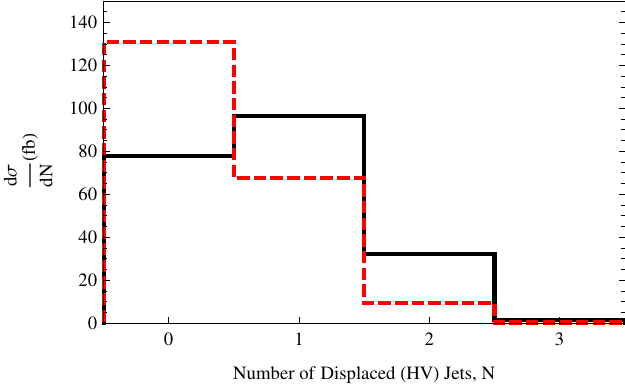}}
  \end{center}
  \caption{The number of displaced objects for the two benchmark points with off-shell hidden sector decays.  Solid black (dashed red) lines represent HV jets (ordinary jets) with displacement greater than 1~mm from the primary vertex.}
  \label{fig:bendisp}
\end{figure}

The last obvious characteristic of our sample points is the presence
or absence of displaced vertices.  We assign a cut on the distance to the
appearance of the first visible  object in a jet or HV jet of
$d > 0.1$~mm to qualify as displaced.
As expected from the discussion
in Section~\ref{subsec:benchP04}, no events from HV2 or HV4 pass
this requirement but a significant number from HV3 and HV5 do.
We plot the total number of objects that qualify as displaced
for HV3 and HV5 in Fig.~\ref{fig:bendisp}.
Note that displaced ``ordinary jets'' are visible decays in the hidden sector
that fail the HV jet criteria but pass our jet conditions.
Comparing these plots to Fig.~\ref{fig:numobj}, we see that in HV3
nearly all HV jets are displaced.  In HV5, we see that when there is only
a single HV jet, it is almost always displaced; only when there
are multiple HV jets are any of them prompt.  This is due to the
Bino decay to $\chi_2^x X_\mu$ dominating over the
$\chi_{1,3}^x X_\mu$ channels.

\begin{figure}[htt]
  \begin{center}
    \includegraphics[width=0.5\textwidth]{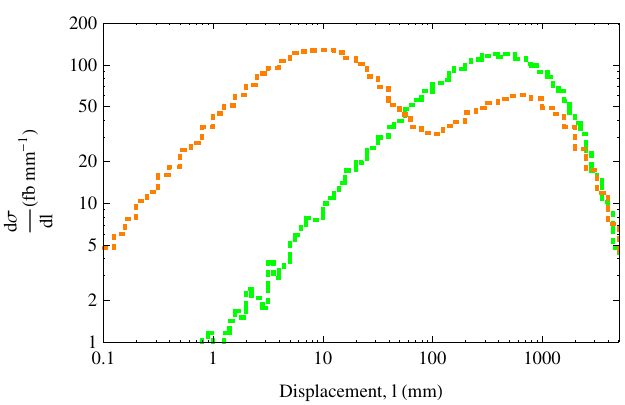}
  \end{center}
  \caption{The distribution of hidden decay lengths from the primary
vertex point for HV3 (green, dot-dashed) and HV5 (orange, dotted).}
  \label{fig:dlength}
\end{figure}

  In Fig.~\ref{fig:dlength}. we plot the net displacements observed in
simulated events for points HV3 and HV5.  No collider resolution effects
have been imposed beyond the lower displacement bound of 0.1~mm.
The distribution for HV3 has a single peak near $d \simeq 40\,\text{cm}$
corresponding to the slow decays $A^x\to h_1^x{X_{\mu}}^*$.
More interestingly, the distribution for HV5 has two distinct peaks
in the distribution of decay lengths.  These correspond to the two
slow decay channels $A^x\to h_1^x{X_{\mu}}^*$
and $\chi_2^x\to\chi_1^x{X_{\mu}}^*$.  Observing such a ``double bump'' feature
could be helpful in distinguishing the hidden sector decay chains.

  So far we have only investigated LHC collider events initiated by gluino
pair production.  Direct electroweak production of neutralinos (and charginos)
could also lead to interesting signals if they produce visible HV jets.
This possibility was investigated in
Refs.~\cite{Baumgart:2009tn,Bai:2009it,Cheung:2009su}, where it was
assumed that the hidden sector produced two HV jets in every LSMP neutralino
decay for a total of four in each supersymmetric event.  In our specific
and explicit realization of a supersymmetric hidden sector, we find that
most supersymmetric events produce many fewer HV jets.  A particularly
distinctive possibility for direct neutralino production that arises
in our scenario are events with HV monojets, consisting of a single
visible HV jet accompanied by missing energy.  Existing LHC monojet
searches~\cite{Aad:2011xw} do not bound our sample
MSSM spectrum since the
production cross section for a 300\,\gev\ Bino-like neutralino
is extremely small ($0.005\,$fb), but such searches with existing
and upcoming data could be relevant for smaller masses or
Wino- or Higgsino-like neutralinos~\cite{Baumgart:2009tn,Cheung:2009su}.

To this end, we show in Table~\ref{tab:Monojet} the fraction of events for each benchmark where pair production of promptly-decaying neutralinos leads to a monojet event.  We use the monojet definition of Ref.~\cite{Aad:2011xw}, specifically one jet (including HV jets) with $p_T > 120$~GeV, no other objects with $p_T > 30$~GeV, and missing energy $\met > 120$~GeV.  HV2 produces such events most frequently, due to producing hard HV jets and having a small $P_2$, see Fig.~\ref{fig:hardHVpt} and Table~\ref{tab:bench}.  HV4 and HV5 tend to produce high-multiplicity final states, while objects in HV3 are soft.  Even so, all four points could be interesting in this channel for an appropriate LSMP.  Distinguishing HV jets from ordinary QCD jets would further extend the reach of such searches.

\begin{table}[ttt]
  \centering
  \begin{tabular}{|c|c|}
    \hline
    Benchmark & Monojet \\
     & Branching Ratio \\
    \hline
    HV1 & 0\% \\
    HV2 & 26\% \\
    HV3 & 11\% \\
    HV4 & 18\% \\
    HV5 & 12\% \\
    \hline
  \end{tabular}
  \caption{Branching ratios for pair production of 300~GeV LSMPs to monojets; see text for details.}
  \label{tab:Monojet}
\end{table}

\bigskip

\section{Conclusions\label{sec:conc}}

  We have presented and investigated the collider signals of a simple
and minimal example of a supersymmetric hidden valley.
The theory consists of the MSSM together with a hidden sector
consisting of a $U(1)_x$ vector multiplet and a pair of chiral Higgs
multiplets.  These two sectors interact with each other exclusively
through supersymmetric gauge kinetic mixing.
We assume that the hidden sector feels supersymmetry breaking
less directly than the visible MSSM sector, and has a characteristic
mass scale (and spontaneous breakdown of $U(1)_x$) near a GeV.

  The light hidden sector can give rise to new and unusual signals at
high energy particle colliders such as the LHC.  MSSM superpartners will
be created in the usual way and decay in a cascade down to the lightest MSSM
superpartner~(LSMP).  In turn, the LSMP will decay into the hidden
sector by way of the kinetic mixing interaction.  The decay cascade
will continue within the hidden sector, and one or more \Ux\
vector bosons can be produced along the way.  Depending on the
hidden spectrum, the vector will frequently decay exclusively
to pairs of light SM particles.  Thus, the continuation of the
supersymmetric cascade into the hidden sector can give rise to
additional \emph{visible} particles in the final state characteristic
of a hidden valley.

  In the present work we have studied the collider signals of the specific
case of a neutralino LSMP.  We find boosted HV jets of collimated leptons or
hadrons, missing energy, and multiple displaced vertices for various parameter
values in the hidden sector.  More general sets of MSSM and hidden sector
parameters could also give rise to heavy-flavour jets, charged track stubs,
long-lived coloured particles, and much else.
The broad range of new and unusual collider signals that can arise
in this simple theory makes it a useful benchmark model with which to
investigate the sensitivity and feasibility of LHC searches for hidden valleys.

\bigskip

\section*{Acknowledgements}

  We thank Anadi Canepa, David Poland, Matthew Strassler,
Oliver Stelzer-Chilton, Itay Yavin, and Kathryn Zurek for helpful discussions.  We also thank an anonymous referee for critical comments that helped improve this work.
DM would like to thank the Perimeter Institute for their hospitality
during the completion of this work.  This research is supported by NSERC.

\appendix

\section{Appendix:  Hidden Sector Interactions\label{appa}}

  We collect here the relevant interactions within the hidden sector,
written in terms of the mass eigenstates.  We work in the unitary gauge,
and our definitions of the mass eigenstates and mixing angles follows
the conventions specified in Section~\ref{subsec:mix}.  We also keep
terms only up to linear order in the small kinetic mixing $\epsilon$.

\subsection{Gauge Interactions}

  The relevant hidden sector fermion gauge couplings are
\bea
-\mathcal{L} &\supset& -i\tilde{H}^{\dagger}\bar{\sigma}^{\mu}
(\del_{\mu}+ig_xX'_{\mu})\tilde{H}
-i{\tilde{H}'}\,\!^{\dagger}\bar{\sigma}^{\mu}
(\del_{\mu}-ig_xX'_{\mu})\tilde{H}'\\
&\supset&
g_x(P_{i1}P_{j1}^*-P_{i2}P_{j2}^*)
{\chi_i^x}^{\dagger}\bar{\sigma}^{\mu}(X'_{\mu}
)\chi_j^x \,,
\nnmb
\eea
where $X'_{\mu} = (X_{\mu}+\epsilon\,s_WZ_{\mu})$.

  For the scalars, we have
\bea
-\mathcal{L}&\supset& -\left|(\del_{\mu}+ig_xX'_{\mu})H\right|^2
-\left|(\del_{\mu}-ig_xX'_{\mu})H'\right|^2
\nnmb\\
&\supset&
g_x(c_{\zeta}R_{1a}-s_{\zeta}R_{2a}){X'}^{\mu}
({A^x}\ccdot\del_{\mu}h_a - \del_{\mu}{A^x}\ccdot h_a)\\
&&
-\frac{1}{2}g_x^2(X'_{\mu})^2(h_1^2+h_2^2+{A^x}^2)
-\sqrt{2}g_x^2(R_{1a}s_{\zeta}+R_{2a}c_{\zeta})\eta(X'_{\mu})^2h_a \,.
\nnmb
\eea
Note that we have dropped various kinetic and mass terms along the way.

\subsection{Yukawa Interactions}

  The relevant terms involving only hidden sector fields are
\bea
-\mathcal{L} &\supset& \sqrt{2}g_xH^*\tilde{H}\tilde{X}
-\sqrt{2}g_x{H'}^*\tilde{H}'\tilde{X} +(h.c.)\nnmb\\
&\supset&
g_x(R_{1a}P^*_{i1}P^*_{j3}-R_{2a}P_{i2}^*P_{j3}^*)h_a\chi_i^x\chi_j^x\\
&&
-ig_x(c_{\zeta}R_{i1}^*R_{j3}^*
-s_{\zeta}P_{i2}^*P_{j3}^*){A^x}\chi_i^x\chi_{j}^x
+(h.c.)\nnmb
\eea
These couplings are the supersymmetric partners of the vector boson
couplings given above.

  There are also interactions with the visible Bino due to the kinetic
mixing shift of Eq.~\eqref{eq:binoxino}.  This leads to
\bea
-\mathcal{L} &\supset& -\epsilon\,\sqrt{2}g_xH^*\tilde{H}\tilde{B}
+\epsilon\,\sqrt{2}g_x{H'}^*\tilde{H}'\tilde{B} +(h.c.)\nnmb\\
&\supset&
-g_x(R_{1a}P^*_{i1}N^*_{j1}-R_{2a}P_{i2}^*N_{j1}^*)h_a\chi_i^x\chi_j^0\\
&&
+ig_x(c_{\zeta}P_{i1}^*N_{j1}^*
-s_{\zeta}P_{i2}^*N_{j1}^*){A^x}\chi_i^x\chi_{j}^0
+(h.c.)\nnmb
\eea
These couplings are crucial to allowing the MSSM LSP to decay to
the lighter hidden sector states.


\subsection{Scalar Interactions}

  These derive from the $D$-term:
\bea
-\mathcal{L} &\supset&
\frac{g_x^2}{2}\left(|H|^2-|H'|^2\right)^2 
\\
&\supset&
\frac{1}{2}g_xm_x\,\left(s_{\zeta}R_{1a}-c_{\zeta}R_{2a}\right)\,
\left[\left(R_{1c}R_{1d}-R_2cR_{2d})\right]\,h_ch_d+c_{2\zeta}{A^x}^2\right]\,h_a
+ \ldots
\nnmb
\eea
In the last line we have omitted various 4-point scalar couplings that are
less important for the collider phenomenology.

\subsection{Mass Mixing with the MSSM}

  Mass mixing between the visible and hidden sectors can largely
be neglected to an excellent approximation.  Such mixing is suppressed
by powers of both $\epsilon$ and $m_x/m_{MSSM}$, leading to very small
mixing angles.  However, as discussed in the text, there are two
situations where hidden-visible mass mixing can play an important role
in the collider signatures of the theory, and we address the issue here.

  Since the relevant mixing angles are very small, we can treat the mass
mixing as a two-point interaction between purely hidden and visible
mass eigenstates.  The first situation of interest is the mass mixing
between the Bino and the hidden Higgsinos arising from
Eq.~\eqref{eq:decaytohid}.  This induces the fermion mass mixing interaction
\beq
-\mathcal{L} \supset -\epsilon\,m_x\,
\left(s_{\zeta}P_{i1}^*N_{j1}^*
-c_{\zeta}P_{i2}^*N_{j1}^*\right)\,\chi_i^x\chi_j^0 \,.
\eeq
By mixing into the hidden Higgsino, the LSMP neutralino develops
a coupling to the hidden vector.  In LSMP decays, the mass-ratio
suppression in the mixing angle is cancelled by the enhanced
coupling to the longitudinal component of the vector.

  The second situation where hidden-visible mass mixing can play an important
role in the phenomenology arises for the lightest $CP$-even hidden Higgs.
Its mass-mixing interaction with the SM Higgs boson derives
from the mixed $D$-term interaction $\epsilon\,D_xD_Y$, and is given by
\bea
-\mathcal{L} &\supset& 
-\epsilon\left(\sum_iY_i|\phi_i|^2\right)\left(\sum_ix_i|\varphi_i|^2\right)
\nnmb\\
&\supset& -\epsilon\,s_W\,m_xm_Z\,
(s_{\zeta}R_{1a}-c_{\zeta}R_{2a})
\label{eq:higgsmix}\\
&&~~~\times \left[
(s_{\beta}\tilde{c}+c_{\beta}\tilde{s})\,h^0
+ (s_{\beta}\tilde{s}-c_{\beta}\tilde{c})\,H^0\right]h_{a}^x \,.
\nnmb
\eea
Even though this coupling leads to a very small mass mixing,
it can still play an important role in the decay of the lightest CP-even
hidden Higgs as we discuss in Section~\ref{subsec:decayout}.

\section{Appendix:  Benchmark Point Parameters\label{appb}}

  We list here the underlying hidden sector parameter points for
the five benchmark points discussed in Section~\ref{sec:bench}.
Recall that $g_x = 0.3$ and $\epsilon = 10^{-3}$ is assumed throughout.

\begin{center}
  \begin{tabular}{|c|c|c|c|c|c|}
    \hline
    & $m_{x}$ (GeV) & $m_{A^x}$ (GeV) & $\mu'$ (GeV) & $M_x$ (GeV) & $\tan\zeta$ \\
    \hline
    HV1 & 5.73 & 0.60 & $-7.47$ & 1.69 & 6.48 \\
    HV2 & 0.65 & 7.43 & $-1.09$ & 0.32 & 0.92 \\
    HV3 & 6.49 & 4.04 & $-2.52$ & 1.21 & 1.85 \\
    HV4 & 0.90 & 9.72 & $-1.83$ & 0.28 & 0.39 \\
    HV5 & 3.53 & 3.00 & $\phantom{-} 3.51$ & 8.41 & 0.39 \\
    \hline
  \end{tabular}
\end{center}

\newpage

\section{Appendix:  Additional Plots \label{appc}}

\begin{figure}[htt]
  \begin{center}
    \subfigure[HV2]{\includegraphics[width=0.47\textwidth]{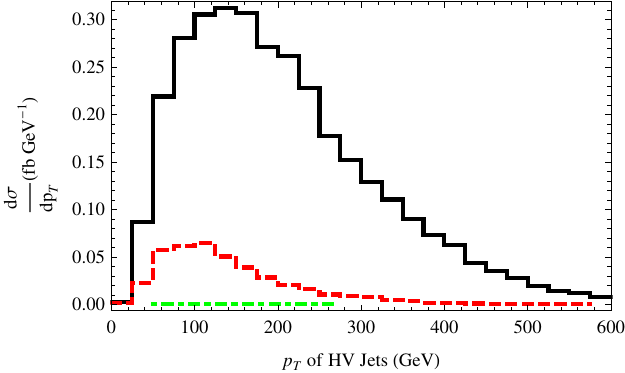}}
    \subfigure[HV3]{\includegraphics[width=0.47\textwidth]{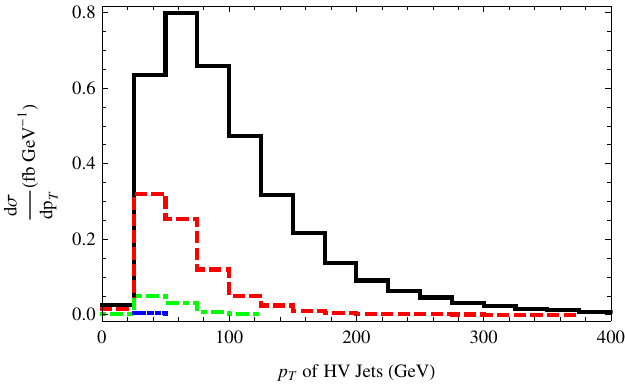}}
    \subfigure[HV4]{\includegraphics[width=0.47\textwidth]{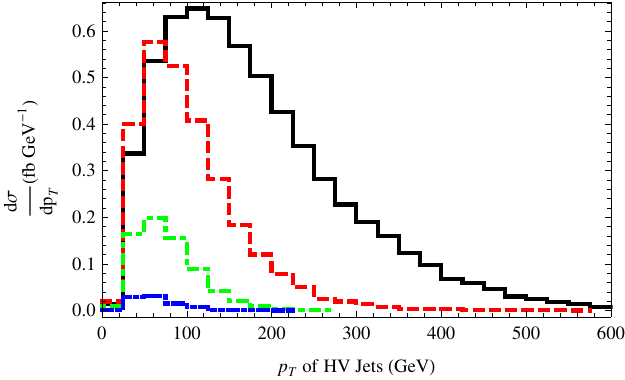}}
    \subfigure[HV5]{\includegraphics[width=0.47\textwidth]{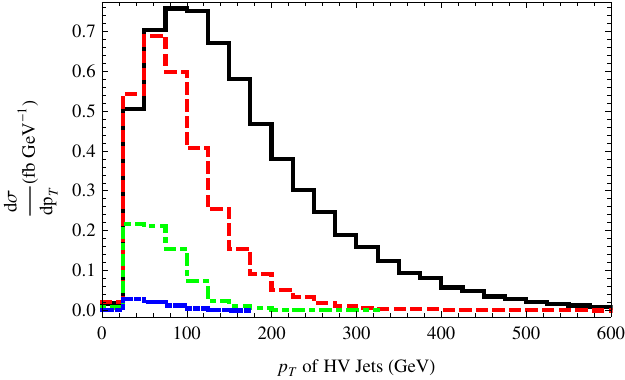}}
  \end{center}
  \caption{The $p_T$ distributions of the HV jets for the four benchmark points 2--5.  The first (second, third, fourth) hardest jets are denoted by solid black (dashed red, dot-dashed green, dot-dot-dashed blue) lines.}
  \label{fig:HVptbenches}
\end{figure}

  We collect here some additional plots related to the discussion
of Section~\ref{sec:coll}.  In Fig.~\ref{fig:HVptbenches} we show
the $p_T$ distributions for the four hardest HV jets for the
sample points HV2--HV5.  Recall that HV1 has no HV jets.

\clearpage

\bibliography{hvbib}{}
\bibliographystyle{JHEP}

\end{document}